\documentclass[a4paper,fleqn]{cas-sc}
\usepackage{setspace}
\usepackage{float}
\usepackage{caption}
\usepackage[authoryear]{natbib}
\usepackage{geometry}
\begin{document}
\let\WriteBookmarks\relax
\def\floatpagepagefraction{1}
\def\textpagefraction{.001}   

% Short title
\shorttitle{}    

% Main title of the paper
\title [mode = title]{Assessing extreme flood impacts on urban rail transit: A passenger-oriented, resilience-informed framework}  

% Author infomation
\author[1,2]{Wei Bi}[orcid=0000-0003-1042-4547]
\cormark[1]

\ead{bi@ibi.baug.ethz.ch}
\ead{wb316@cam.ac.uk}

\affiliation[1]{
    organization={Centre for Sustainable Development, Department of Engineering, University of Cambridge},
    city={Cambridge CB2 1PZ},
    country={United Kingdom}
}

\affiliation[2]{
    organization={Institute of Construction and Infrastructure Management, Department of Civil, Environmental and Geomatic Engineering, ETH Zurich},
    city={8093 Zurich},
    country={Switzerland}
}

\cortext[cor1]{Corresponding author}

%******************************** Abstract ************************************

\begin{abstract}
Urban rail transit systems (URTSs) are increasingly exposed to extreme floods following heavy precipitation, yet passenger travel impacts are often assessed through delay-based indicators that overlook infeasible journeys under large-scale disruptions. This study develops a passenger-oriented, resilience-informed framework for assessing flood impacts on URTS journeys from disruption onset to recovery completion. The framework presents a novel six-category classification of journey impacts, explicitly considering rerouting, alternative station use, and a delay threshold. It is demonstrated through hourly dynamic simulations of 15 London URTS lines under 30-year, 100-year, and 1,000-year flood risk scenarios. Results indicate that severe flood disruptions lead to substantial unsatisfied demand, driven primarily by unavailable routes rather than unacceptable delays. Compared with finer behaviour adjustments, rerouting dominates travel impacts. These findings highlight the significance of moving beyond delay-based assessment and provide valuable evidence on essential behavioural mechanisms for strategic-level stress testing intended to inform URTS flood resilience intervention planning. 
\end{abstract}

%******************************** keywords ************************************


\maketitle
% \doublespacing
\onehalfspacing

%******************************** Introduction ************************************
\section{Introduction}
\label{section:1}

Urban rail transit systems (URTSs), including metros, light rail, and trams, are crucial for underpinning sustainable mobility, equitable access, and economic activity, carrying over 58 billion passenger journeys in 2023 and continuing to expand globally \citep{UITP2025Global}. Alongside the growth, URTSs are facing increasing challenges from extreme weather events, particularly heavy rainfall, as parts of their infrastructure are directly exposed to flash flooding \citep{Koks2019global,Liu2023Global}. Climate projections of more frequent and intense precipitation at global and continental scales \citep{Seneviratne2023Weather} suggest that such events are unlikely to remain exceptional, which highlights the need for approaches capable of anticipating the consequences of plausible but unusual disruption scenarios rather than relying solely on historical averages.

Resilience analysis has emerged as a central paradigm for assessing the impacts of low-probability but high-consequence disruptions on infrastructure systems \citep{Linkov2014Changing}. An essential step for a resilience-informed approach is to stress-test the system for generating disruption scenarios. Stress testing employs “what-if” analysis to examine the effects of “extreme but plausible” disruptions that may not be captured in routine risk assessments \citep{UNECE2024Stress,Linkov2022Resilience}. By modelling a range of adverse scenarios and monitoring how the system performs under these conditions, stress testing helps diagnose critical weak points, assess system resilience to different levels of risk, and inform adaptation planning to mitigate future impacts. 

A meaningful resilience-informed approach therefore requires impact indicators that can translate simulated disruptions into measurable changes in system performance. These indicators can be defined across multiple dimensions \citep{Bi2023Old,Bruneau2003Framework}, reflecting the sociotechnical nature of URTSs and their role in supporting urban functioning and broader socioeconomic activity. The physical dimension considers infrastructure asset damage (reviewed in \cite{Bai2025Resilience}) and the resulting loss of network connectivity, with the latter being the dominant focus of network-based transport resilience studies and typically indicated by topological attributes such as network global efficiency \citep{Chen2024Resilience,Guo2025How,Hao2023Improving,Hu2025Surrogatebased,Hu2024Postearthquake,Martello2021Evaluation,Zhang2018Resiliency,Zhu2025Resilience}, betweenness centrality \citep{Bhatta2024Dynamics,DaSilvaFradique2025Enhanced,Jiao2024Research,Lu2024Traffic}, the average shortest path length \citep{Bhatta2024Dynamics,Chopra2016networkbased,He2024Functionality,Qi2021Resilience}, and the giant connected components \citep{Guo2025How,Ma2022Measuring,Yadav2020Resilience,Zhu2025Resilience}. Despite their widespread use, topology-based indicators offer a limited representation of disruption impacts, as they describe the structural properties of the network but do not capture how service operations or passenger journeys are affected in practice. The social dimension examines how disruption affects passengers' ability to reach essential opportunities, such as employment, healthcare, and education, and how these impacts vary across population groups \citep{Bi2025Supporting}. Equity considerations have become increasingly popular in transport resilience research \citep{Coleman2024Weaving}, recognising that service disruptions may impose uneven burdens depending on passengers' mobility needs, socioeconomic circumstances, and access to alternative transport \citep{TransportforLondon2024Equity,SorianoA.2022Inclusive}. Within this dimension, current analyses are primarily built on accessibility-based indicators that quantify changes in destination reachability under disrupted network conditions \citep{Boakye2022role,Liu2025HumanCentric,vanMarle2023Including}. The economic dimension captures the financial consequences of disruption \citep{Rose2007Economic}, including direct costs associated with infrastructure damage, emergency response, and lost operating revenue \citep{Bi2024Assessing,Nasrazadani2025SimulationBased}, as well as indirect losses arising from passenger travel time delays, supply-chain and logistics inefficiencies, and broader productivity losses across the urban economy \citep{Sikka2013Whata,Tatano2008framework,Wei2022datadriven,Wei2018Disaster}. 

This study examines URTS flood disruption impacts from an operational dimension, with a focus on passenger travel outcomes. Operational performance reflects passenger travel outcomes during service disruptions and is therefore directly relevant to transport operators \citep{NationalAcademiesofSciencesEngineeringandMedicine2021Investing}, whose decisions on recovery prioritisation, emergency crew deployment, and contingency service planning shape how disruptions unfold and are managed. Operational performance analysis also bridges physical network disruptions and broader socioeconomic consequences: physical damage only translates into welfare and economic losses through the degradation of service, and the magnitude of those downstream losses is largely determined by how operations respond to and recover from the shock. Within this dimension, travel delay-based indicators have been widely used to quantify performance loss from transport disruptions, as they provide an operationally meaningful representation of how passenger journeys are affected. The specific formulations differ in how journey time is measured or decomposed, including delay minutes \citep{Adjetey-Bahun2016model,Ilalokhoin2023model,Li2026Disentangling,Li2026Stresstesting}, average travel time \citep{DaSilvaFradique2025Enhanced}, total travel time \citep{Faturechi2014Travela}, waiting time \citep{Zhang2022Characterizing}, and travel time reliability \citep{Chepuri2018Examininga,Sikka2013Whata}. However, delay-based indicators are conditional on journey feasibility, as delay can only be recorded for journeys that can still be completed through the disrupted system. This limits their applicability for measuring the impacts of large-scale URTS disruption scenarios, such as those triggered by extreme weather events, where a considerable share of planned journeys cannot be completed within the disrupted system and therefore falls outside a delay-based account of disruption impact.

Measuring unsatisfied travel demand therefore provides a useful complement for capturing large-scale disruption impacts and has been adopted in recent URTS resilience studies \citep{Goldbeck2019Resilience,Liu2024Enhancing,Zhao2022Evaluating}. Despite this, few studies have developed a detailed framework for classifying the impacts of URTS flood disruptions on passenger travel into logically differentiated categories. Such a framework should account not only for delay and unsatisfied travel demand, but also for passenger behavioural responses to disruptions, which play a critical role in shaping travel outcomes. \cite{Zhao2022Evaluating} categorise the impacts of metro station closure on passenger travel outcomes by distinguishing between fulfilled and unfulfilled trips, and further differentiating completed trips according to changes in origin–destination choice, route choice, queuing time, and overall travel time. While this classification provides a useful basis, further work is needed to develop a more explicit hierarchical framework that examine how key behavioural assumptions, such as rerouting choices, delay tolerance, and alternative station availability, affect the resulting impact estimates. Moreover, their assessment does not explicitly model the recovery process and therefore does not fully capture how passenger travel impacts evolve from disruption onset through to recovery completion. 

To address (1) the absence of a detailed framework for classifying URTS flood disruption impacts on passenger travel into logically differentiated categories, and (2) the lack of evidence on the relative importance of passenger behavioural mechanisms in shaping aggregate disruption impacts from disruption onset to recovery completion, this study proposes a passenger-oriented, resilience-informed framework for assessing extreme flood impacts on URTS journeys. The framework extends the author’s previously published model for assessing URTS flood resilience \citep{Bi2024Assessing} by incorporating three passenger behavioural mechanisms into an hourly dynamic simulation of flood-induced disruption and recovery on a hourly basis: rerouting along alternative paths within the rail network, the use of alternative origin and/or destination stations within walking distance, and an acceptable-delay threshold beyond which a journey is treated as unsatisfied. Through this integration, the impact on every planned journey is decomposed into one of six categories, including journeys that follow the original shortest path (A); journeys that reroute between the original origin and destination stations within the delay threshold (B) ; journeys completed through alternative origin and/or destination stations within a walking distance with (C) or without delay (D); journeys that cannot be completed because no available path or alternative origin and/or destination stations (E), and  journeys that cannot be completed because the minimum achievable delay exceeds the threshold (F). An empirical demonstration is conducted using the London URTS, which comprises five modes across 15 lines and serves around 5.3 million daily passenger journeys through 443 stations. Three surface water flood risk scenarios are examined, covering a moderate-to-extreme range with return periods of 1 in 30, 1 in 100, and 1 in 1,000 years. This study contributes to the field of large-scale, passenger-oriented transport resilience assessments in the following ways:

\begin{enumerate}[(1)]
    \item This study demonstrates the value of moving beyond delay-based passenger impact assessment for extreme flood disruptions. Where substantial numbers of journeys cannot be completed, delay alone provides only a partial account of disruption severity. The framework therefore supports a finer basis for assessing passenger-oriented disruption impacts and for evaluating the effectiveness of resilience interventions, such as asset-level flood protection and alternative recovery strategies, according to how they affect different journey outcomes. These insights offer evidence to support flood resilience appraisal and investment planning for urban rail systems. 
    \item The results provide evidence on the level of behavioural detail required in strategic flood stress testing. The London URTS case demonstrates that rerouting is the dominant behavioural mechanism shaping passenger travel impacts, with its influence substantially outweighing that of alternative station use or delay-based behavioural thresholds. This provides a valuable basis for targeted model simplification in system-level strategic applications in practice, where behavioural features with limited influence on passenger travel outcomes may be simplified to improve computational efficiency for testing a wider range of flood risk scenarios. 
\end{enumerate}

The remainder of this paper is structured as follows. Section \ref{section:2} provides an overview of the URTS flood resilience assessment model. Section \ref{section:3} presents the proposed passenger-oriented impact framework and its integration with the resilience assessment model through a time-stepped simulation procedure. Section \ref{section:4} introduces the London URTSs case study. Section \ref{section:5} presents the results, while Section \ref{section:6} briefly discusses their implications. Section \ref{section:7} outlines the main conclusions and limitations.

%******************************** Overview of the URTS flood resilience assessment model ************************************
\section{Overview of the URTS flood resilience assessment model}
\label{section:2}

This study employs a system-level, resilience-based approach to evaluate how flood-induced URTS service disruptions affect passenger travel. This section outlines the URTS flood resilience assessment model that supports the application of the proposed passenger-oriented impact framework in Section \ref{section:3}). The model was developed and presented in full in the author's recent work \citep{Bi2024Assessing}, and is summarised here only to the level of detail required to support the proposed framework and interpret the case study results. The model is based on the widely applied resilience triangle curve proposed by \cite{Bruneau2003Framework}, where system resilience is indicated by the total performance loss throughout the disruption period. To operationalise this conceptual curve, the model employs a stress testing approach, simulating how URTS operations are affected by flooding and how the system subsequently recovers.

\subsection{Constructing a URTS network model}
\label{section:2.1}

\begin{figure}[ht!] 
    \centering    
    \includegraphics[width=1.0\textwidth]{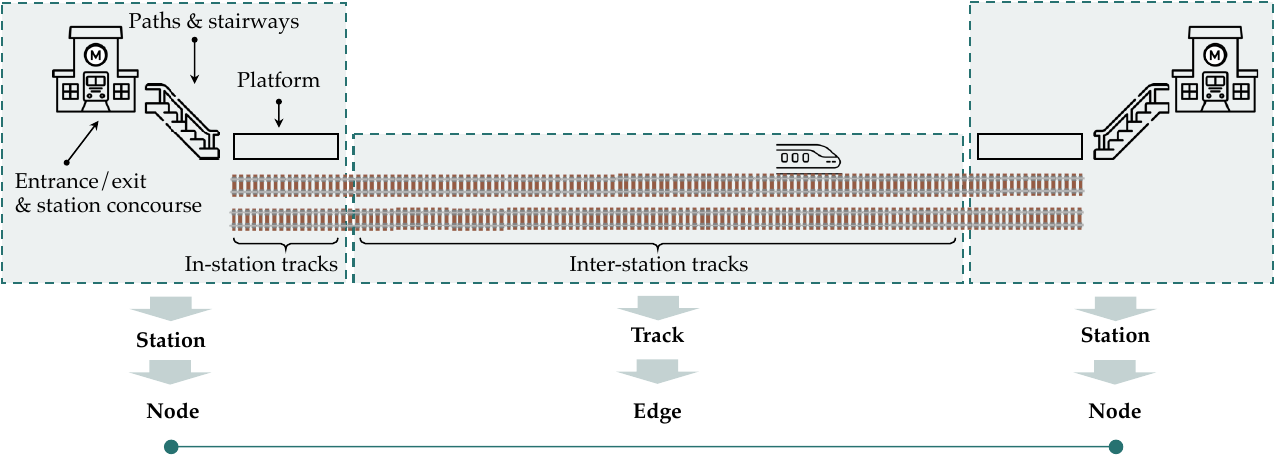}
    \caption {URTS components and their representation in network models \citep{Bi2024Assessing}}
    \label{fig:1}
\end{figure}

A network model $G=(V,E,W)$ is constructed to represent the URTS. $V = \{v_i \mid i =1,2,\ldots\,n\}$ denotes the node set, where each node corresponds to a station and captures its key internal facilities (see Fig. \ref{fig:1}). $E=\{e_{ij}=(v_i,v_j) \,|\, i,j=1,2,\ldots,n; i\neq j\}$ defines the edge set, with each edge representing the track connection between a pair of neighbouring stations. $W=\{w_{ij} \,|\, i,j=1,2,\ldots,n; i\neq j\}$ refers to edge weights, which are specified as the travel times along the corresponding track segments. The connectivity of the network is represented by an adjacency matrix $A$, with entries $A_{ij}(1 \leq i,j \leq n)$ defined as follows: 

\begin{equation}
    A_{ij} = \left\{
    \begin{array}{ll}
    w_{ij}, & \text{if } v_i \text{ and } v_j \text{ are directly connected} \\
    0, & \text{if } v_i \text{ and } v_j \text{ are not directly connected}
    \end{array}
    \right.
    \label{eq:1}
\end{equation}

\noindent where $A_{ij}=w_{ij}$ if stations $v_i$ and $v_j$ are directly linked by a track segment $(v_i,v_j)$, with $w_{ij}$ representing the associated travel time; otherwise, $A_{ij}=0$. $G$ is an undirected graph, allowing bidirectional travel between $v_i$ and $v_j$. This network model is employed to simulate service disruptions in Section \ref{section:2.2} and subsequent recovery processes in Section \ref{section:2.3} for impact analysis.

\subsection{Generating flood disruption scenarios}
\label{section:2.2}

This study utilises flood depth maps to determine URTS stations and track segments that are susceptible to flood risk and consequent loss of functionality when floods occur. The procedure first overlays geographic locations of potential flood entry points with flood depth maps to assess the exposure of stations and track segments and to estimate the corresponding inundation depths. Four categories of flood entry points are considered: (1) station entrances and (2) in-station tracks, which are used to detect station flooding and corresponding node failures; and (3) inter-station tracks and (4) tunnel entrances, which are used to detect track flooding and corresponding edge failures. Flood depth thresholds are then defined to specify the inundation levels at which individual network elements are assumed to become non-operational. Additionally, elements experiencing underground flooding are flagged, as their restoration is typically more time-consuming and subject to a narrower set of feasible recovery options (elaborated later in Section \ref{section:2.3}). The resulting disruption information is then incorporated into the network model to simulate the operational impacts on the URTS, following the details provided in Table \ref{table:1}.

\begin{table}[ht!]
    \footnotesize 
    \caption{Simulating flood disruption effects on URTSs through network modelling (adapted from \cite{Bi2024Assessing})}
    \label{table:1}
    \renewcommand{\arraystretch}{1.2}
    \newcolumntype{L}[1]{>{\raggedright\arraybackslash}p{#1}}
    \begin{tabular}{L{0.07\textwidth} L{0.1\textwidth} L{0.28\textwidth} L{0.44\textwidth}}
        \hline
        {Flooding type} & {Flooding location} & {Disruption effects in practical operation} & {Corresponding simulation in network modelling} \\ \hline
        Station flooding & Station concourse only & The station is closed to passenger boarding and alighting, while trains are still able to travel through the station without stopping. & The node denoting the affected station is excluded from the network model. To allow trains to continue passing through, a temporary node and the relevant adjacent edges are added at the same location. \\
         & Track within stations & The station is closed, and trains cannot pass. & The node denoting the affected station is excluded from the network model. \\
        Track flooding & Track between stations & 
        Train services cannot operate along the flooded track segment. Disruption of an inter-station track may also limit access to nearby turning points or substations, thereby affecting the operation of adjacent track sections and potentially causing partial suspension of the line. & The edge corresponding to the flooded track segment is excluded from the network model. In addition, the two neighbouring edges on either side of the affected track segment are suspended to provide a simplified representation of cascading disruption effects.  \\ \hline
    \end{tabular}
\end{table}

\subsection{Modelling recovery processes}
\label{section:2.3}

Modelling URTS recovery from flood disruptions is a key component of a resilience-based approach, distinguishing it from other analysis such as risk assessments. It requires three types of critical information: (1) feasible methods for cleaning up floodwater, (2) the recovery time required for stations and tracks at varying flood depths, and (3) the scheduling priorities for allocating critical yet limited recovery resources to flooded elements.

\noindent \textbf{(1) Recovery option}

Table \ref{table:2} summarises feasible recovery options for cleaning up floodwater. The most efficient option is to allocate an emergency crew to pump out water, which is applicable to all situations. Nevertheless, only a limited number of emergency crews can operate simultaneously. In cases where emergency crews are not immediately available, manual water removal by station staff using basic equipment, such as mops and buckets, provides an alternative. It is assumed that stations have sufficient staff capacity for this task. For health and safety reasons, however, manual removal is not considered suitable for underground flooding or for cases where water levels exceed the threshold for safe access. Finally, in cases where floodwater affects surface-level tracks located in an open environment, it is possible to allow the water to recede naturally, although this is not feasible for recovering from underground flooding. Where more than one recovery option is feasible and available, they are prioritised according to their relative efficiency.

\begin{table}[ht!]
    \footnotesize 
    \caption{Recovery options}
    \label{table:2}
    \renewcommand{\arraystretch}{1.2}
    \newcolumntype{L}[1]{>{\raggedright\arraybackslash}p{#1}}
    \begin{tabular}{L{0.16\textwidth} L{0.32\textwidth} L{0.22\textwidth} L{0.19\textwidth}}
        \hline
        {Recovery option} & {Applicability} & {Resource availability} & {Efficiency \& priority} \\ \hline
        Emergency crew pumps out floodwater & Any situation. & Limited number of emergency crews. & $\star\star\star$ \\
        Manual water removal by station staff & Any situation except: \newline - Underground flooding. \newline - Where the flood depth exceeds the safe-access threshold adopted in this study, set at 0.6 m. & Stations have sufficient staff capacity for this task. & $\star\star$ \\
        Naturally recede & Any situation except: \newline - Underground flooding. & No resource required. & $\star$ \\ \hline
    \end{tabular}
\end{table}

\noindent \textbf{(2) Recovery times}

For an equivalent flood severity, the recovery times are ranked as $t_p < t_m < t_n$, where $t_p$, $t_m$, and $t_n$ correspond to the times for recovery by pumping, manual cleanup, and natural drainage, respectively. The estimation of these recovery times also needs to consider flood depths, which can reflect the extent of flood damage: deeper floods generally require longer recovery periods \citep{Martello2023Depth}. In the case study, these estimations can be reasonably made based on available records of past flood incidents. 

Moreover, the recovery time for an element can be dynamic, influenced by the timing as well as the type of recovery option deployed. For instance, if a station is disrupted by shallow flooding and is assigned a medium priority in the resource scheduling process, station staff may begin manual cleanup before an emergency crew becomes available and subsequently takes over using pumping equipment. In this situation, the overall recovery time $T$ dedicated to this station is:

\begin{equation}
    T = t_0 + \left(1 - \frac{t_0}{t_m}\right) t_p
    \label{eq:2}
\end{equation}

\noindent where $t_0$ refers to the initial time used for manual cleanup by station staff before the arrival of an emergency crew, $t_m$ is the duration needed if the station were restored using manual cleanup only, and $t_p$ is the time required if pumping were used throughout. It should be noted that recovery time considers only the time required to clear floodwater. Catastrophic flood damage requiring reconstruction is not accounted for in the recovery modelling due to data limitations.

\noindent \textbf{(3) Resource scheduling priorities}

As emergency crews, the most efficient restoration resource, are often limited in availability, prioritising which flooded elements receive them is crucial for mitigating post-flood impacts. Prioritising the recovery of the most critical elements of the URTS allows for the maximum restoration of services in the shortest possible time. This study adopts a deterministic approach, employing demand-weighted betweenness centrality to assess the importance of flooded elements, which is then used to determine their priorities for receiving emergency crews \citep{Bi2024Assessing}. This demand-weighted betweenness centrality is a combination of topological importance and operational importance. Specifically, betweenness centrality indicates the extent to which a node locates at shortest paths connecting other node pairs. To incorporate the operational dimension into this topological metric, the travel demand between each origin-destination node pair, $d(s,t)$, is introduced as a weight, so that heavily used shortest paths contribute more to the centrality measure:

\begin{equation}
    g(v) = \frac{1}{D} \left( \sum_{\substack{s,t \in V \\ s \neq t \neq v}} \frac{\sigma(s,t|v)d(s,t)}{\sigma(s,t)} \right)
    \label{eq:3}
\end{equation}

\noindent where $g(v)$ refers to the demand-weighted betweenness centrality of node $v$; $\sigma(s,t|v)$ denotes the number of shortest paths between nodes $s$ and $t$ that travel across node $v$; $\sigma(s,t)$ indicates the total number of shortest paths connecting nodes $s$ and $t$; and $D$ denotes the travel demand across all node pairs. The importance of a flooded edge is assigned as the larger centrality value of its two endpoint nodes.

%******************************** Proposed passenger-oriented impact framework ************************************
\section{Proposed passenger-oriented impact framework}
\label{section:3}

\begin{figure}[ht!] 
    \centering    
    \includegraphics[width=1.0\textwidth]{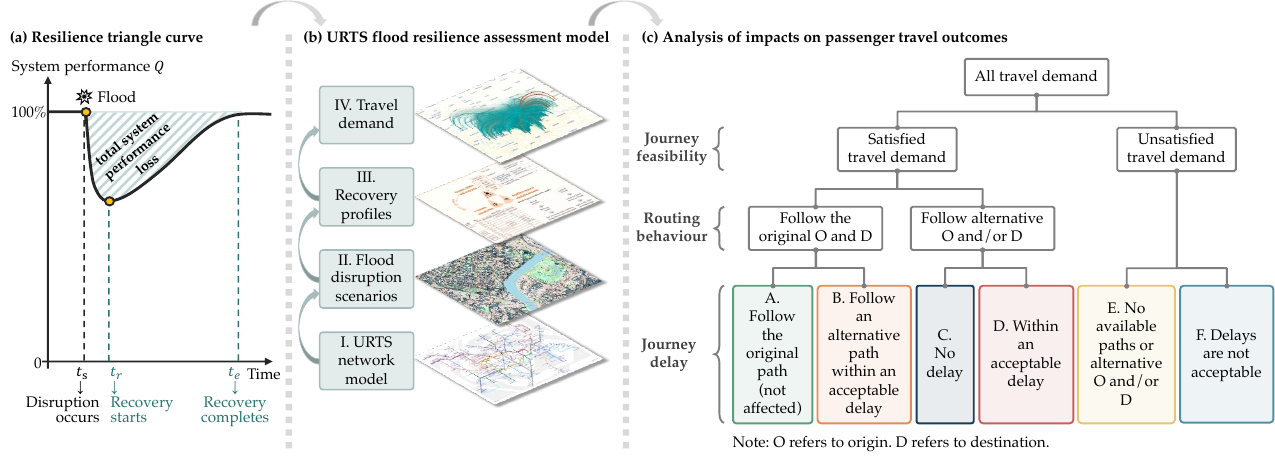}
    \caption {Research framework}
    \label{fig:2}
\end{figure}

Building on the modelling capacity described in Section \ref{section:2}, this section proposes a passenger-oriented impact framework, which evaluates how URTS travel demand is dynamically affected from the onset of a flood-induced service disruption through to the completion of recovery. This framework decomposes the disruption impact on passenger travel according to the behavioural mechanism through which each journey is either completed or unsatisfied, yielding six mutually exclusive impact categories. The overall research framework of this study, integrating the theoretical foundation of the resilience triangle curve, the employed URTS flood resilience assessment model, and the proposed passenger-oriented impact framework, is presented in Fig. \ref{fig:2}. 

\subsection{Consideration of passenger behaviour change}
\label{section:3.1}

This study assumes that passengers travel through the shortest path, defined as the path with the minimum travel time that considers in-vehicle time, transfer time, and station dwell time. While passenger route choice in practice is influenced by additional factors such as crowding, familiarity, and individual preferences, the shortest-path assumption is a common and widely adopted simplification in transport studies \citep{Adjetey-Bahun2016model,Goldbeck2019Resilience,Liu2022Cascadingb}. When service disruptions occur due to flooding, passengers are allowed to reroute within the rail transit network to complete their journeys when alternative paths are available. 

However, not all alternative paths that remain topologically available in the disrupted rail transit network are behaviourally feasible. When rerouting produces excessive delay, passengers may switch to other modes or cancel their trips rather than endure the extended journey. Journey delay therefore should be considered a second determinant of passenger routing behaviour under disruption to distinguish feasible from infeasible rerouting. 

Taken together, this study assumes that a planned journey is satisfied only when both of the following requirements are fulfilled: (1) the origin (O) and destination (D) stations remain connected by at least one feasible path; and (2) where the feasible path with the shortest travel time is not the same as the original shortest path, the additional delay associated with this rerouted journey does not exceed 30 minutes. This 30-minute delay threshold is estimated based on the refund standard for journey delays set by \cite{TransportforLondon2023Londonb,TransportforLondon2023TubeDLRDelays}, and its validity is examined in the results. Journeys that do not meet both conditions are considered unsatisfied. While some such journeys would in practice be accommodated by other modes, including cars, buses, or taxis, they still represent a performance loss to the URTS and are therefore treated as unsatisfied URTS demand. 

\begin{figure}[ht!] 
    \centering    
    \includegraphics[width=0.95\textwidth]{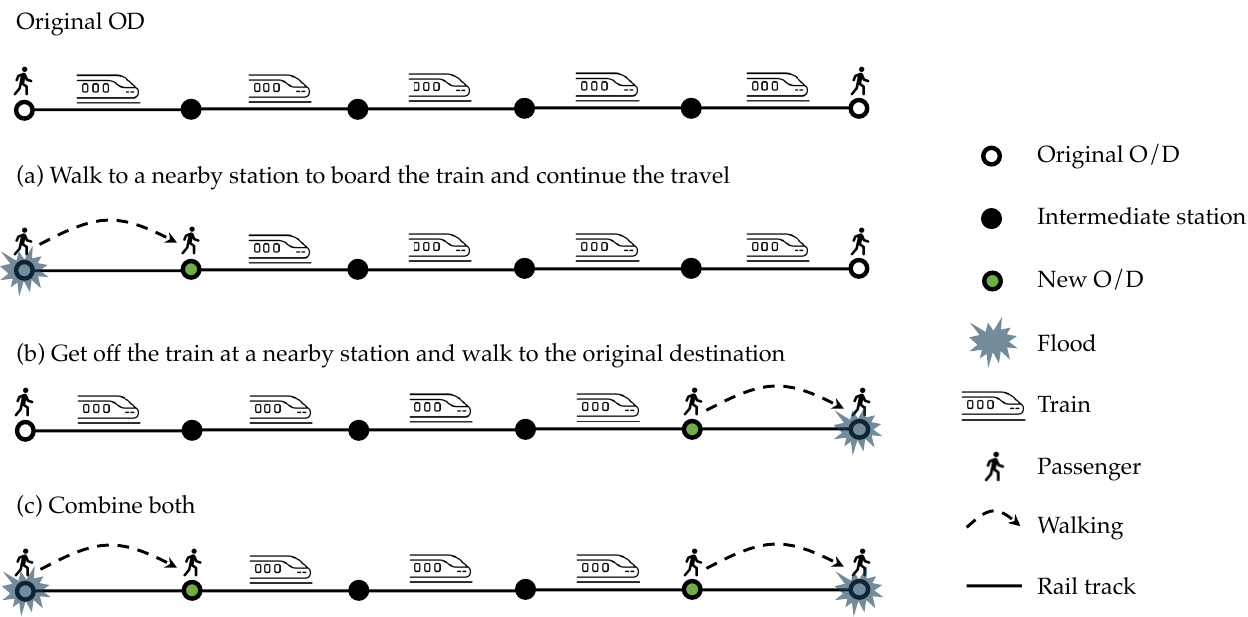}
    \caption {Scenarios of the alternative origin (O) and destination (D) stations}
    \label{fig:3}
\end{figure}

Another consideration regarding changes in passenger behaviour is that, since few public transport trips are completed without any walking, it is reasonable to expect that passengers may walk to a nearby station to continue their journey when their original origin and/or destination station is flooded. Fig. \ref{fig:3} depicts three possible scenarios: (1) when the original origin station is flooded, passengers walk to a nearby station to board a train and continue their journey; (2) when the original destination station is flooded, passengers alight at a nearby station and then walk to their destination; and (3) a combination of both. A walking distance of 800 m is generally considered as a standard threshold, corresponding to approximately 10 minutes of walking time \citep{CIHT2015Planning, Gunn2017Identifying}. In the situation where flood disruptions occur, passengers may be willing to walk farther to complete their journeys, as fewer alternatives are likely to be available. Accordingly, this study permits passengers to walk to a nearby station within a 1 km walking distance, following the assumption adopted in \cite{Zhao2022Evaluating}. This walking time is considered a compromise that passengers are willing to make to complete their journey. It is therefore excluded from the total journey time, while the additional delay associated with the new route is still required to satisfy the 30-minute threshold.

\subsection{Categorisation of disruption impacts on passenger travel}
\label{section:3.2}

As presented in Fig. \ref{fig:2}(c), the decomposition of URTS service disruption impacts on passenger travel is developed through a three-level hierarchy applied to every planned journey. The top level concerns \textit{journey feasibility}, separating satisfied from unsatisfied demand according to whether at least one feasible path exists in the disrupted network within the acceptable delay. The middle level considers \textit{routing behaviour}, which only applies to satisfied journeys and distinguishes those completed between the original origin and destination from those that rely on an alternative O/D within walking distance. This distinction is treated as a separate level because alternative-station use and path rerouting represent behaviourally distinct adaptations to disruption. Separating them makes the role of alternative-station use explicit in the impact analysis. The bottom level captures \textit{journey delay}, recording whether a satisfied journey incurs no delay or a delay within the acceptable threshold, and whether an unsatisfied journey fails because no feasible path or alternative O/D exists or because the minimum achievable delay exceeds the threshold. 

The three levels together yield six categories that span all possible passenger travel outcomes through a URTS under flood disruption. As introduced, satisfied journeys are classified as: (A) those completed following the original origin and destination stations and the original path (i.e., unaffected); (B) those completed following the original origin and destination stations but an alternative path within the delay threshold; (C) those completed through alternative origin and/or destination stations with no delay; and (D) those completed through alternative origin and/or destination stations within the delay threshold. Unsatisfied journeys are classified as resulting from either (E) unavailable paths or alternative origin and/or destination stations, or (F) unacceptable delays. 

\begin{figure}[ht!] 
    \centering    
    \includegraphics[width=1.0\textwidth]{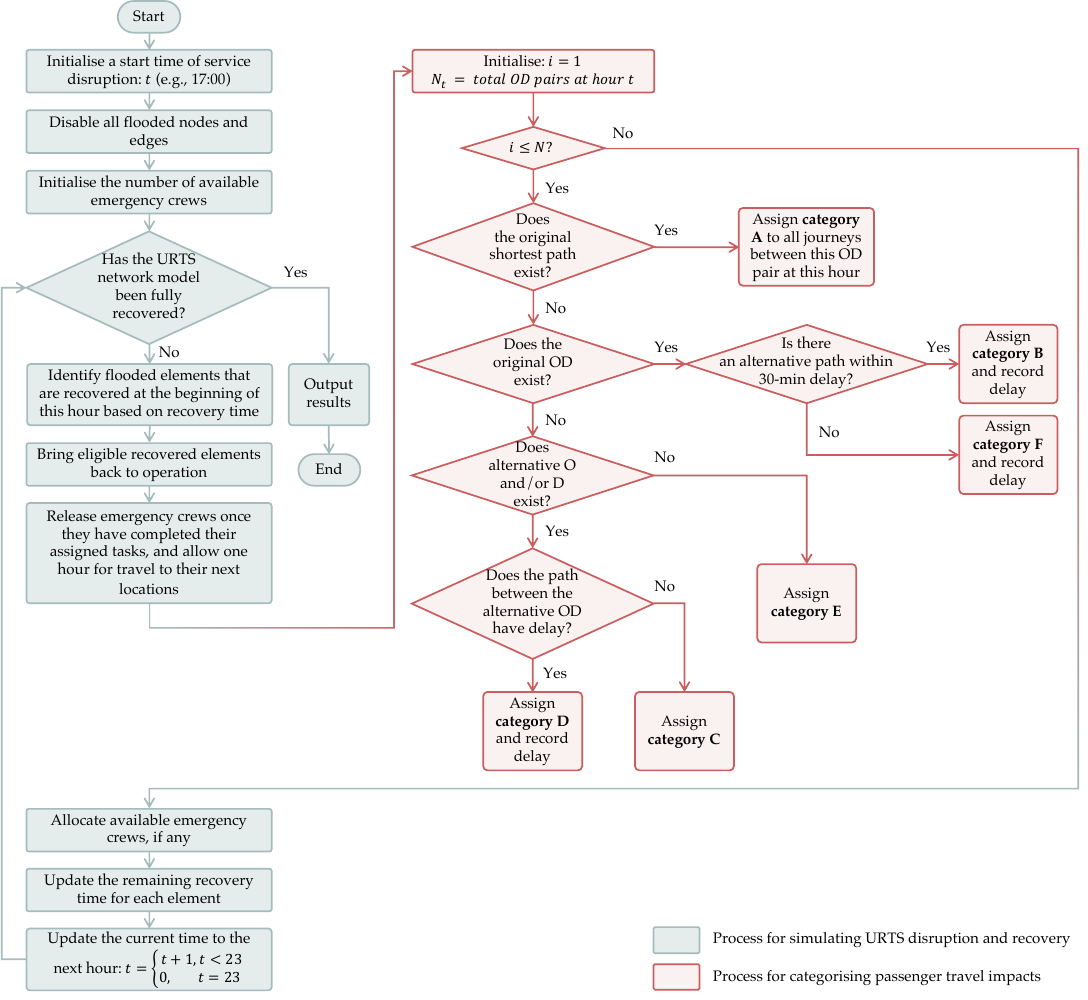}
    \caption {Simulation process of URTS disruption, recovery, and passenger travel impacts}
    \label{fig:4}
\end{figure}

To classify each planned journey according to the impact it experiences from the onset of a flood-induced disruption through to the completion of recovery, the behavioural mechanism described in Section \ref{section:3.1} is integrated into the URTS flood resilience model. The resulting simulation procedure is summarised in Fig. \ref{fig:4}. At the disruption onset, every node and edge identified flooded is disabled in the network model and the available emergency crews are initialised. The simulation then iterates hour by hour: elements whose recovery has completed at the beginning of this hour are set to return to service; every OD pair with travel demand at this hour is processed through a decision sequence and its journeys are assigned to one of the six categories A–F; emergency crews that have finished their previous assignment and completed the one-hour travel to a new location are re-allocated to the remaining flooded elements at the end of this hour, the remaining recovery times required for each flooded element are updated, and the simulation proceeds to the next hour. The simulation terminates once the network has fully recovered, at which point aggregated results across the full disruption and recovery cycle are output.

%************************** Case and data descriptions ******************************
\section{Case and data descriptions}
\label{section:4}

\subsection{London multi-modal rail transit system}
\label{section:4.1}

\begin{figure}[ht!] 
    \centering    
    \includegraphics[width=0.95\textwidth]{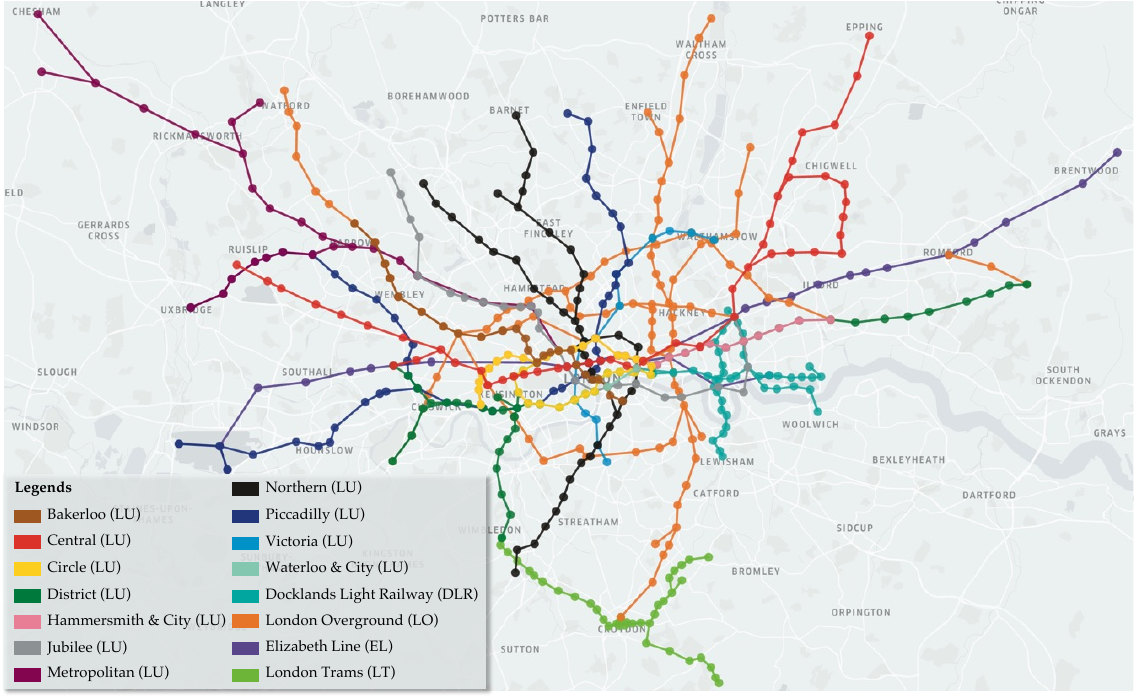}
    \caption {15 lines of the London URTS (adapted from \cite{Bi2024Assessing})}
    \label{fig:5}
\end{figure}

This study selects the London URTS as a case study, given its high network complexity and substantial exposure to surface water flood risk \citep{GreaterLondonAuthority2018London}. It is operated by the experienced transit agency, Transport for London (TfL), which provides valuable public data. As presented in Fig. \ref{fig:5}, fifteen lines of the London URTS are investigated, including eleven lines of the London Underground (LU), six routes of the London Overground, the Elizabeth Line, the Dockland Light Railway, and the London Trams. The network model built for the London URTS comprises 443 nodes and 533 edges. The travel time on each edge is derived from TfL timetables \citep{TransportforLondon2023ElizabethTimetables,TransportforLondon2023OvergroundTimetables,TransportforLondon2023UndergroundTimetables,TransportforLondon2023TramTimeTable}. The average station transfer time is obtained from TfL \citep{TransportforLondon2011InterchangeWalkingTimes}. The network model-based simulation of disruption and recovery is implemented in Python using NetworkX.

\subsection{Hourly travel demand}
\label{secton:4.2}

\begin{figure}[ht!] 
    \centering    
    \includegraphics[width=1.0\textwidth]{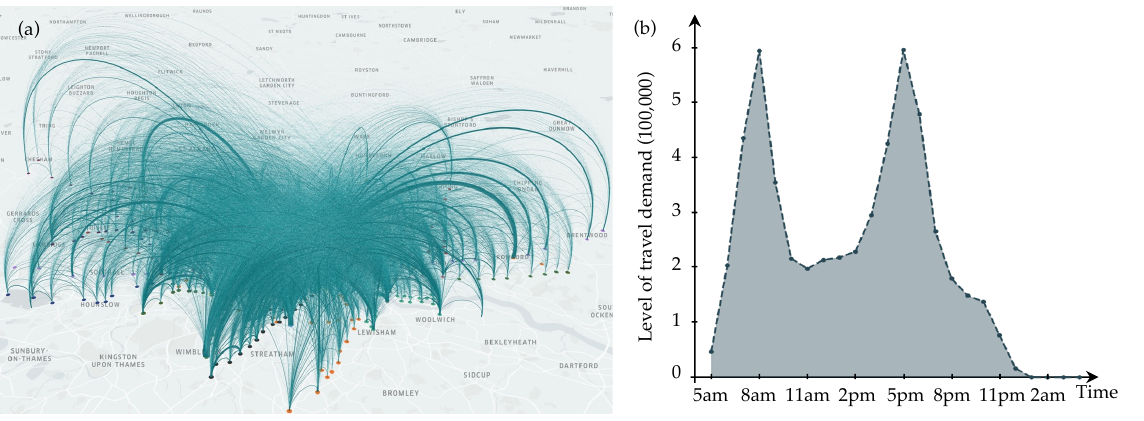}
    \caption {Weekday travel demand through the London URTS in 2019}
    \label{fig:6}
\end{figure}

The London URTS accommodates the daily travel needs of millions under normal operation conditions. This study utilises the data of travel demand on a typical weekday in 2019, prior to the Covid-19 pandemic, comprising 5,342,647 trips across 51,905 origin-destination station pairs, as presented in Fig. \ref{fig:6}(a). This dataset is publicly available from \cite{TransportforLondon2020CrowdingData} and provides journey statistics for any OD pairs in 15-minute intervals. In this study, these records are consolidated into hourly time periods for analysis (see Fig. \ref{fig:6}(b)).

\subsection{Alternative origin and destination stations}
\label{section:4.3}

\begin{figure}[ht!] 
    \centering    
    \includegraphics[width=0.95\textwidth]{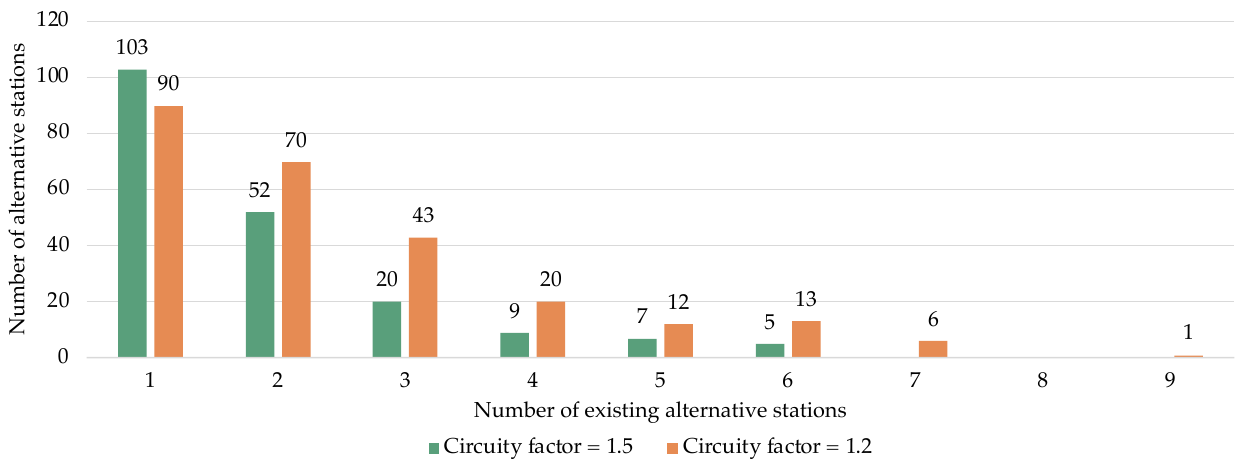}
    \caption {Number of existing alternative stations for each station}
    \label{fig:7}
\end{figure}

To identify stations within a 1 km walking distance of each other, a circuity factor is applied to account for the difference between the straight-line distance and the actual walking distance along the street network, ensuring a simplified yet accurate estimation of pedestrian route lengths. The circuity factors of 1.5 (i.e., the straight-line distance is around 667 m) and 1.2 (i.e., the straight-line distance is around 833 m) are selected based on references \cite{Ballou2002Selected} and \cite{Meeder2018Measuring}. When setting the circuity factor as 1.5, 196 out of 443 stations of the London URTS are identified as having at least one alternative station within walking distance. Setting the circuity factor to 1.2 increases this number to 255 stations. The statistics for the number of alternative stations for each station are presented in Fig. \ref{fig:7}. Compared with a circuity factor of 1.5, a circuity factor of 1.2 increases the number of alternatives available per station. In both cases, stations with only one or two alternatives account for the largest share, while the number of stations declines markedly as the number of alternatives increases. This suggests that walk-based flexibility is available for a substantial proportion of stations, but is generally limited to a small number of nearby alternatives. 

\subsection{Surface water flood disruption scenarios}
\label{section:4.4}

Three scenarios are investigated, including the 30-year, 100-year, and 1,000-year flood disruption scenarios from \cite{Bi2024Assessing}. These scenarios were developed from UK official surface water flood depth maps \citep{DepartmentofEnvironmentFoodRuralAffairs2021Defra}, which classify inundation depth into six intervals using threshold values of 0.15 m, 0.3 m, 0.6 m, 0.9 m, and 1.2 m. With a horizontal resolution of 2 m, these official maps enable flood exposure to be assessed at the level of individual URTS flood entry points.

\begin{figure}[ht!] 
    \centering    
    \includegraphics[width=1.0\textwidth]{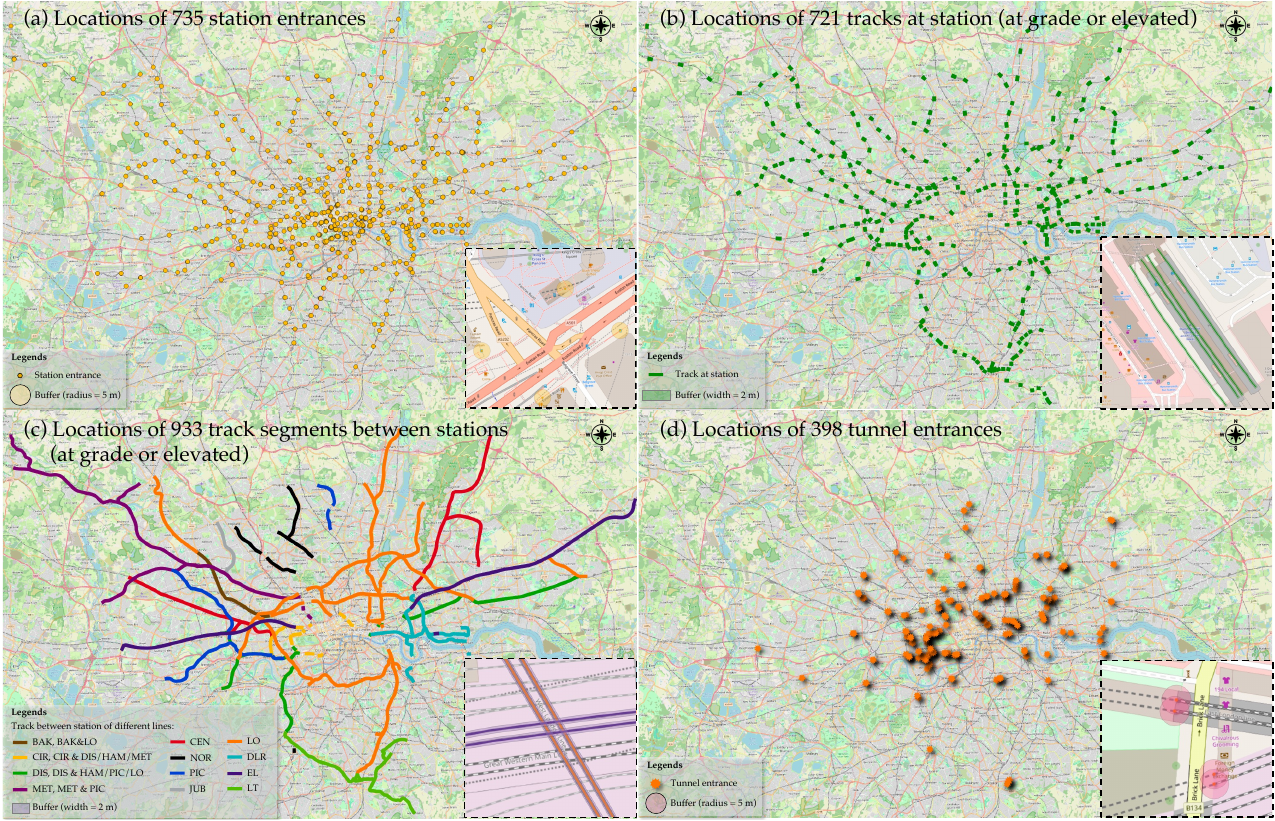}
    \caption {Location of examined flood entry points of the London URTS}
    \label{fig:8}
\end{figure}

The London URTS analysis considers 735 station entrances, 721 in-station track sections, 933 inter-station track sections, and 398 tunnel entrances to determine their corresponding flood depths (see Fig. \ref{fig:8}). A 0.3 m failure threshold is then applied to non-underground station entrances, in-station track sections, tunnel entrances, and inter-station track sections. Regarding entrances of underground stations, a higher depth threshold of 0.6 m is set. At this level, water exceeding 0.6 m at the entrance can cascade down staircases into the underground area, disrupting operations. Following this, stations (represented as nodes) and tracks (represented as edges) that become dysfunctional due to flooding can be identified, as shown in Fig. \ref{fig:B.1} in \ref{app:B}. The number of flooded elements within each depth interval for each scenario is detailed in Table \ref{table:B.1}.

\subsection{Recovery profiles of the London URTS}
\label{section:4.5}

Consultations with TfL professionals suggest that, for large-scale flood disruptions affecting the London URTS, it is reasonable to assume the simultaneous deployment of up to 15 emergency crews \citep{Bi2024Assessing}. Priority will be given to flooded elements with higher scheduling priority, calculated based on Eq. (\ref{eq:3}). 

The estimated recovery duration for each flooded node or edge varies between 3 and 48 hours, depending on the inundation depth, whether the flooding occurs in an underground space, and the recovery method used. The full set of recovery-time assumptions is introduced in Table \ref{table:A.1} in \ref{app:A}. These assumptions are based on discussions with TfL professionals and reference to their historical flood incident records.

%******************************** Results ************************************
\section{Results}
\label{section:5}

\begin{figure}[ht!] 
    \centering    
    \includegraphics[width=1.0\textwidth]{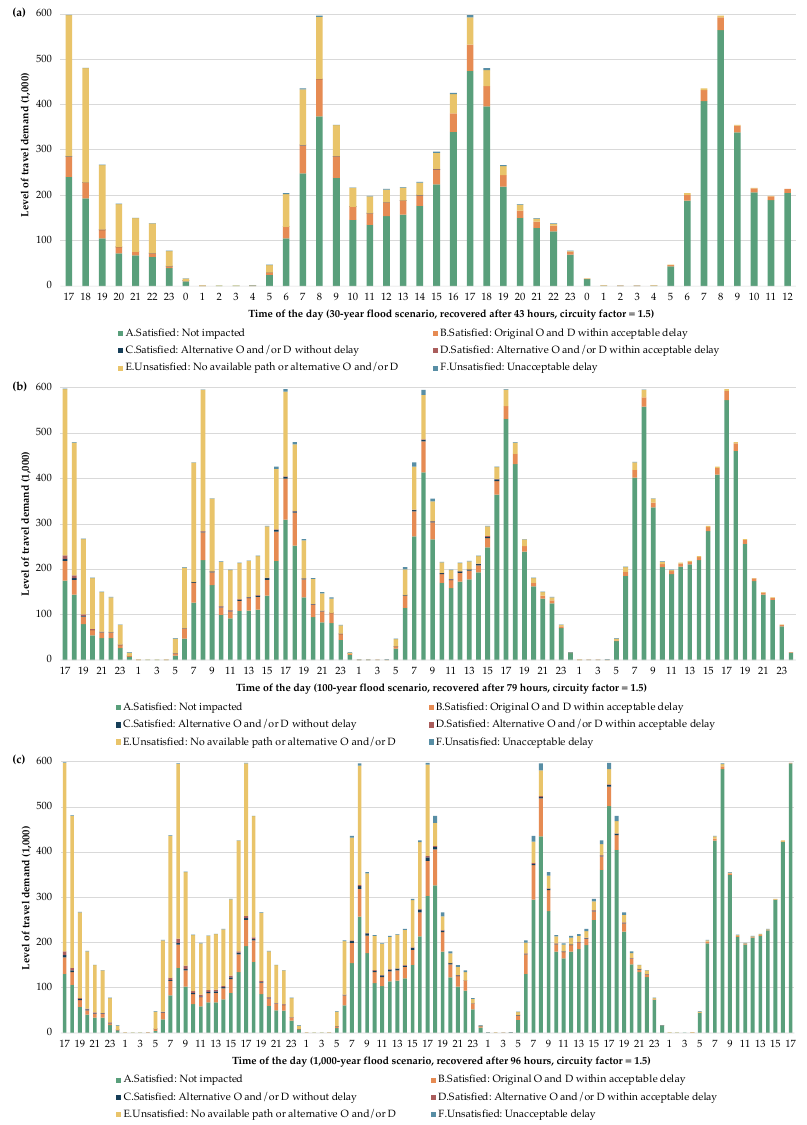}
    \caption {Hourly distribution of travel demand across impact categories (A) – (F) (circuity factor = 1.5)}
    \label{fig:9}
\end{figure}

Following the simulation process in Fig. \ref{fig:4}, the service disruption start time is set at 17:00, aligning with the late afternoon timing of two flood events that disrupted the London URTS in July 2021. Fig. \ref{fig:9} presents the hourly distribution of travel demand by impact category under each flood risk scenario when the circuity factor is set to 1.5, with each bar representing the total level of travel demand in a given hour and each coloured segment indicating the number of journeys classified into categories (A) – (F). The corresponding results for a circuity factor of 1.2 are provided in Fig. \ref{fig:C.1} in \ref{app:C}. Table \ref{table:4} provides the statistical summary of the number of journeys classified into each category across flood risk scenarios. 

\begin{table}[ht!]
    \footnotesize
    \caption{The number of journeys classified into impact categories (A) – (F)}
    \label{table:4}
    \renewcommand{\arraystretch}{1.2}
    \begin{tabular}{@{}p{0.08\linewidth}p{0.06\linewidth}p{0.1\linewidth}p{0.06\linewidth}p{0.06\linewidth}p{0.06\linewidth}p{0.06\linewidth}p{0.06\linewidth}p{0.06\linewidth}p{0.07\linewidth}p{0.07\linewidth}@{}}
    \toprule
    {Flood scenario} & \multicolumn{6}{c}{{Satisfied travel demand (1,000)}} & \multicolumn{4}{c}{{Unsatisfied travel demand (1,000)}} \\
    \cmidrule(lr){2-7} \cmidrule(lr){8-11}
    & {A. Not affected} & {B. Follow an alternative path between the original OD within an acceptable delay} & \multicolumn{2}{>{\raggedright\arraybackslash}p{0.1\linewidth}}{C. Follow an alternative OD with no delay} & \multicolumn{2}{>{\raggedright\arraybackslash}p{0.1\linewidth}}{D. Follow an alternative OD within an acceptable delay} & \multicolumn{2}{>{\raggedright\arraybackslash}p{0.1\linewidth}}{E. No available paths or alternative O and/or D} & \multicolumn{2}{>{\raggedright\arraybackslash}p{0.1\linewidth}}{F. Delays are not acceptable} \\
    \cmidrule(lr){4-5} \cmidrule(lr){6-7} \cmidrule(lr){8-9} \cmidrule(lr){10-11}
     & & & {CF=1.5} & {CF=1.2} & {CF=1.5} & {CF=1.2} & {CF=1.5} & {CF=1.2} & {CF=1.5} & {CF=1.2} \\
    \midrule
    30-year & 6,832 & 849 & 4 & 9 & 5 & 29 & 1,794 & 1,756 & 31 & 40 \\
    100-year & 12,400 & 1,403 & 77 & 89 & 40 & 93 & 3,938 & 3,858 & 74 & 89 \\
    1,000-year & 13,012 & 1,798 & 158 & 187 & 124 & 198 & 6,688 & 6,542 & 187 & 230 \\
    \bottomrule
    \end{tabular}

\vspace{0.5\baselineskip}
\small\raggedright Note: CF is the short for circuity factor.
\end{table}

\subsection{High levels of unsatisfied travel demand due to unavailable routes}
\label{section:5.1}

\begin{figure}[ht!] 
    \centering    
    \includegraphics[width=1.0\textwidth]{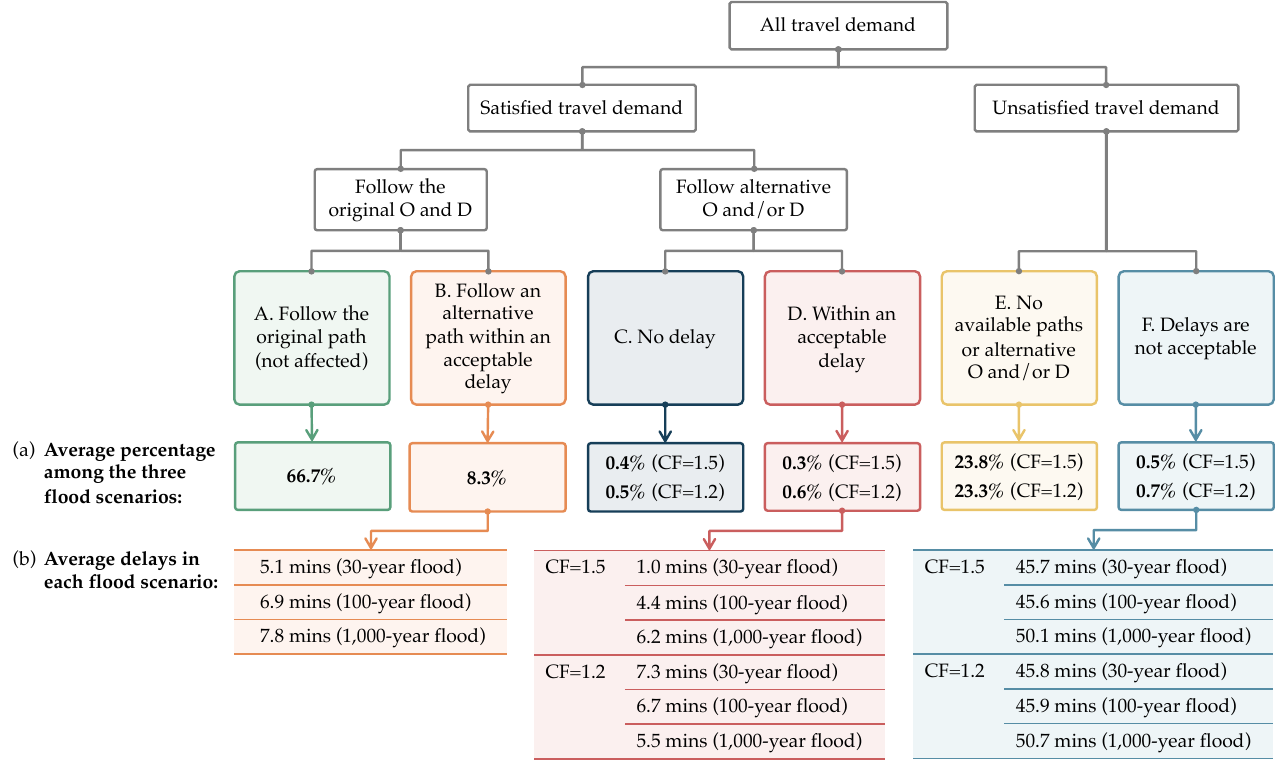}
    \caption {Average journey shares by category and average delays across the three flood risk scenarios}
    \label{fig:10}
\end{figure}

Fig. \ref{fig:10} demonstrates the significant impacts of flood-induced service disruptions on passenger travel, particularly in the initial stages, when over half of the travel demand is disrupted and remains unsatisfied (details in \ref{app:D}). The majority of unsatisfied travel demand is due to the absence of available paths or alternative origin and/or destination stations (category E), which accounts for approximately 98\% across all three flood risk scenarios and both circuity factors. Only 2\% of the total unsatisfied travel demand is attributed to unacceptable delays of the alternative paths (category F). 

These results highlight two implications for resilience-based approaches to assessing transport service disruption impacts. First, in plausible large-scale flood disruption scenarios, delay alone is not sufficient to characterise passenger impacts, as its relevance diminishes when a substantial proportion of journeys cannot be completed. In such cases, unsatisfied travel demand should be included as a critical complementary indicator to capture disruption severity more comprehensively. Second, although behavioural change in response to excessive delay is often considered an important component of transport disruption modelling, the results indicate that its contribution to additional unmet demand is limited across the scenarios examined. This suggests that, for severe and widespread disruption events, modelling emphasis may be more appropriately placed on the loss of feasible routes than on delay-induced behavioural thresholds, and that a simplified measure of the latter may be acceptable where computational efficiency is a concern.

\subsection{Varying impacts on satisfied travel demand}
\label{section:5.2}

A detailed examination of how individual passenger journeys are completed shows that the majority of satisfied travel demand is unaffected by service disruptions (category A), as the original shortest paths remain available. On average, these journeys account for 66.7\% of total travel demand over the disruption periods across the three flood risk scenarios. This high average share is partly explained by the later stages of disruption, when the proportion of unaffected journeys increases as services are progressively restored. This is evident in Fig. \ref{fig:9}, where category A gradually becomes more dominant over time. Nevertheless, even during the initial stages of disruption and the early stages of post-flood recovery, unaffected journeys still constitute the majority of satisfied travel demand, as detailed in \ref{app:C}. Travel demand satisfied through alternative paths between the original origin and destination stations constitutes the second largest share (category B), averaging 8.3\% across the three flood scenarios (see Fig. \ref{fig:10}(a)). This share remains substantially lower than that of unaffected satisfied demand and declines at later stages, particularly on the final day of disruption, as services are progressively restored. Travel demand satisfied through walking to alternative stations (categories C and D) is negligible in the 30-year flood risk scenario and becomes more apparent in the 100-year and 1,000-year flood scenarios. Within these journeys, those associated with acceptable delays (category D) exceed those with no delay (category C) at the beginning of recovery, whereas the reverse is observed during the intermediate recovery stage as more services resume. Setting the circuity factor to 1.2 yields similar results, as indicated in Table \ref{table:4}, although this allows for a slightly greater amount of travel demand to be satisfied, amounting to 29K, 65K, and 103K in the 30-year, 100-year, and 1,000-year flood risk scenarios, respectively. 

These results indicate that the maintenance of satisfied travel demand during flood disruption depends primarily on the continued availability of original shortest paths, rather than on extensive passenger rerouting or station substitution. This highlights the inherent robustness of the London URTS, while also indicating that passenger adaptation within the network tends to play a secondary, rather than dominant, role in maintaining overall travel demand satisfaction. Furthermore, the relatively modest effect of accounting for alternative stations, which represents only around 1\% of total travel demand (see Fig. \ref{fig:10}(a)), suggests that this feature may be simplified in transport disruption modelling where computational efficiency is a priority. Nevertheless, the slightly greater role of alternative stations in the more severe flood risk scenarios suggests that behavioural flexibility remains relevant, particularly when flooding affects station access more extensively.

\subsection{Average travel delays}
\label{section:5.3}

As shown in Fig. \ref{fig:10}(b), under the assumption of a 30-minute acceptable delay threshold, the average delays for journeys completed via original OD pair but alternative paths are 5.1 minutes, 6.9 minutes, and 7.8 minutes in the three flood scenarios, respectively. These values remain well below the threshold, indicating that the disruption impact on such journeys is generally modest. The average delays in journeys involving alternative origin and/or destination stations are also minimal, remaining under 10 minutes across various scenarios and circuity factors. As introduced in Section \ref{section:3.2}, this travel time delay does not consider any additional walking time associated with access to alternative stations. Even with an additional 20 minutes of walking taken into account, comprising 10 minutes to an alternative origin station and a further 10 minutes from the destination-side station
(as illustrated in Fig. \ref{fig:3}(c)), the overall delay stays within the 30-minute threshold. Finally, the estimated average delays in unsatisfied travel demand resulting from excessive delays are also evaluated, ranging between 45 and 51 minutes. Delays of this magnitude are generally unlikely to be acceptable to passengers. These findings support the use of a 30-minute acceptable delay threshold as a reasonable assumption.

%******************************** Discussion *************************************
\section{Discussion}
\label{section:6}

\subsection{Beyond delay-based passenger impact assessment}
\label{section:6.1}

The results show that delay alone does not adequately capture passenger impacts under severe flood disruption. In the examined scenarios, the dominant source of unsatisfied travel demand is not excessive delay on available services, but the absence of feasible routes. Under such conditions, an assessment framework centred primarily on delay or travel time reliability is liable to underestimate disruption impacts. This finding is important for the appraisal of transport adaptation measures. A conventional delay-based evaluation is well suited to identifying interventions that reduce travel delay costs on remaining services, but is less effective in representing measures whose main benefit lies in preserving journey feasibility. In large-scale disruption scenarios, an intervention may produce only a modest reduction in average delay but generate substantial passenger benefit by shifting journeys from unsatisfied to satisfied categories. The presented passenger impact category framework makes this distinction explicit by separating fundamentally different travel outcomes, including unaffected journeys, delayed but completed journeys, and journeys that cannot be completed at all. This provides a more discriminating basis for assessing flood impacts on passenger travel, particularly for large-scale disruptions where there is considerable loss of network accessibility rather than just modest deterioration in travel time performance.

\subsection{Behavioural simplification in URTS flood stress testing at strategic level}
\label{section:6.2}

The results also inform the behavioural change details essential for URTS flood stress testing at strategic level. In the London case, compared to rerouting, alternative origin and destination station use has very limited contribution to satisfying journeys, and unmet demand caused by unacceptable delay accounts for only a small share of total disrupted journeys. This indicates that, for strategic-level flood stress testing intended to inform resilience intervention planning, a selective approach to behavioural representation may be justified. This is particularly relevant when dozens of scenarios need to be evaluated, making the analysis computationally demanding. In such contexts, it is reasonable and practical to prioritise the modelling of rerouting behaviour, while representing finer behavioural adjustments, such as alternative station use or delay thresholds, in a more simplified manner to save computational resources. That said, it should be note that this is less appropriate for analysis that aim to predict individual passenger responses in detail, where richer behavioural representation is more critical, such as transport agent-based modelling.

%******************************** Conclusions ************************************
\section{Conclusions}
\label{section:7}

URTSs are expected to face increasing risk of large-scale flood disruption due to more frequent and intense extreme rainfall events under climate change, which gives rise to a growing need for resilience and adaptation planning supported by stress testing plausible scenarios and quantifying specific disruption impacts on passenger travel. This study presents a passenger-oriented, resilience-informed framework for assessing extreme flood impacts on URTS journeys. Drawing on an established model assessing URTS flood resilience, the framework extends disruption analysis from operational performance loss to detailed passenger travel outcomes. Specifically, it distinguishes between satisfied and unsatisfied travel demand and further categorises journeys according to whether they remain unaffected (category A), require rerouting between the original OD pairs (category B), rely on alternative origin and destination stations (categories C and D), or become impossible to complete as a result of unavailable routes (category E) or unacceptable delays (category F). Applied to the 15 rail transit lines in London under 30-year, 100-year, and 1,000-year surface water flood risk scenarios, the framework enables a dynamic assessment of how passenger impacts evolve throughout disruption and recovery. 

The case study results show that flood-induced service disruptions can lead to substantial passenger impacts, particularly in the early stages of disruption, when over 50\% of travel demand is unsatisfied. Across the three scenarios, the dominant source of unsatisfied travel demand (approximately 98\%) is the absence of available paths or accessible origin and destination stations, whereas only a minimal proportion is attributable to unacceptable delay on alternative routes. Among satisfied journeys, most remain unaffected because original shortest paths continue to be available, while rerouted journeys constitute the second largest share. Journeys completed through alternative origin and destination stations are present but minimal. These findings indicate that, in severe flood disruption scenarios, passenger travel impacts are shaped primarily by journey infeasibility rather than by travel delay alone. 

Based on these results, this study highlights that assessing passenger impacts, particularly under large-scale disruption scenarios, requires going beyond delay-based metrics to account explicitly for journeys that become impossible to complete. It also demonstrates that, for system-level strategic flood stress testing, behavioural change detail may be represented selectively, since rerouting has a much greater influence on passenger travel impacts than alternative station use or delay-based behavioural thresholds in the London case. Finally, the proposed framework could be extended to evaluate the effectiveness of specific resilience interventions, such as asset-level flood protection and alternative recovery strategies, in order to compare changes in the composition of passenger travel impacts. More broadly, the methodology provides a basis for integrating passenger-oriented impact assessment into URTS flood resilience appraisal.

Several limitations should be noted. Firstly, the flood disruption scenarios are derived from static flood depth maps and therefore do not capture the temporal evolution of flooding during an event. As a result, the analysis does not capture how changing flood conditions over time may affect URTS operations from flood onset to peak disruption. Future research could address this by integrating dynamic flood simulation with URTS disruption modelling. Secondly, while the model captures interactions among different urban rail transit modes, including mass transit, light rail, and tram services, it does not represent interactions with other road transport modes such as buses or taxis. Within the modelling boundary adopted here, all unsatisfied travel demand is treated as a performance loss of the rail transit system itself. In practice, however, some of these journeys may still be completed by alternative modes, which could affect the estimated passenger travel outcomes when viewed at the wider transport system level. Future research could extend the framework to incorporate such intermodal interactions, though this would require a more computationally intensive modelling approach, additional data on road-based transport services and traffic conditions, and simulation of flood-induced disruption on the road network itself.

%********************************* Appendix *************************************

\appendix
\renewcommand{\thesection}{Appendix \Alph{section}}

%******************************** Appendix A ************************************

\newpage
\section{Recovery time assumptions for flooded nodes and edges}
\label{app:A}

\renewcommand{\thetable}{A.\arabic{table}} 
\setcounter{table}{0}

\begin{table}[ht!]
    \footnotesize
    \caption{Element recovery time adopted in \cite{Bi2024Assessing}}
    \label{table:A.1}
    \renewcommand{\arraystretch}{1.15}
    \begin{tabular}{@{}p{0.07\linewidth}p{0.22\linewidth}p{0.1\linewidth}p{0.15\linewidth}p{0.15\linewidth}p{0.18\linewidth}@{}}
    \toprule
    \multirow[t]{2}{0.07\textwidth}{Element type} & {Flood location} & \multirow[t]{2}{0.1\textwidth}{Flood depth (m)} & \multicolumn{3}{c}{{Recovery time (h)}} \\
    \cmidrule(lr){4-6}
     & & & {An emergency crew pumps out water $t_p$} & {Manual water removal by station staff $t_m$} & {Naturally recede $t_n$}\\
    \midrule
    \multirow[t]{7}{*}{Node} & \multirow[t]{4}{3.5cm}{Flooding at station concourse and/or surface-level in-station tracks} & 0.3-0.6 & 3 & 6 & -(manually cleared) \\
     & & 0.6-0.9 & 9 & - & 22 \\
     & & 0.9-1.2 & 15 & - & 36 \\
     & & $>$1.2 & 20 & - & 48 \\
     & \multirow[t]{3}{3.5cm}{Flooding at concourse and underground in-station tracks} & 0.6-0.9 & 14 & -(pump only) & -(pump only) \\
     & & 0.9-1.2 & 22 & -(pump only) & -(pump only) \\
     & & $>$1.2 & 30 & -(pump only) & -(pump only) \\
    \multirow[t]{8}{*}{Edge} & \multirow[t]{4}{3.5cm}{Flooding at surface-level inter-station tracks at open environments} & 0.3-0.6 & 3 & 6 & -(manually cleared) \\
     & & 0.6-0.9 & 9 & - & 22 \\
     & & 0.9-1.2 & 15 & - & 36 \\
     & & $>$1.2 & 20 & - & 48 \\
     & \multirow[t]{4}{3.5cm}{Flooding at underground inter-station tracks within tunnel sections} & 0.3-0.6 & 2* & -(pump only) & -(pump only) \\
     & & 0.6-0.9 & 5* & -(pump only) & -(pump only) \\
	 & & 0.9-1.2 & 7* & -(pump only) & -(pump only) \\
	 & & $>$1.2 & 10* & -(pump only) & -(pump only) \\
    \bottomrule
    \end{tabular}

\vspace{0.5\baselineskip}
\small\raggedright Note: * indicates additional recovery time for tunnel flooding and should be added to the recovery time for the corresponding open-section inter-station track section. For instance, if an inter-station track experiences 0.9–1.2 m flooding in its open section and 0.6–0.9 m flooding in its tunnel section, the total recovery time using pumps is 15 + 5 = 20 hours.
\end{table}

%******************************** Appendix B ************************************

\newpage
\section{London URTS flood disruption scenarios}
\label{app:B}

\renewcommand{\thefigure}{B.\arabic{figure}}
\setcounter{figure}{0}

\begin{figure}[ht!] 
    \centering    
    \includegraphics[width=1.0\textwidth]{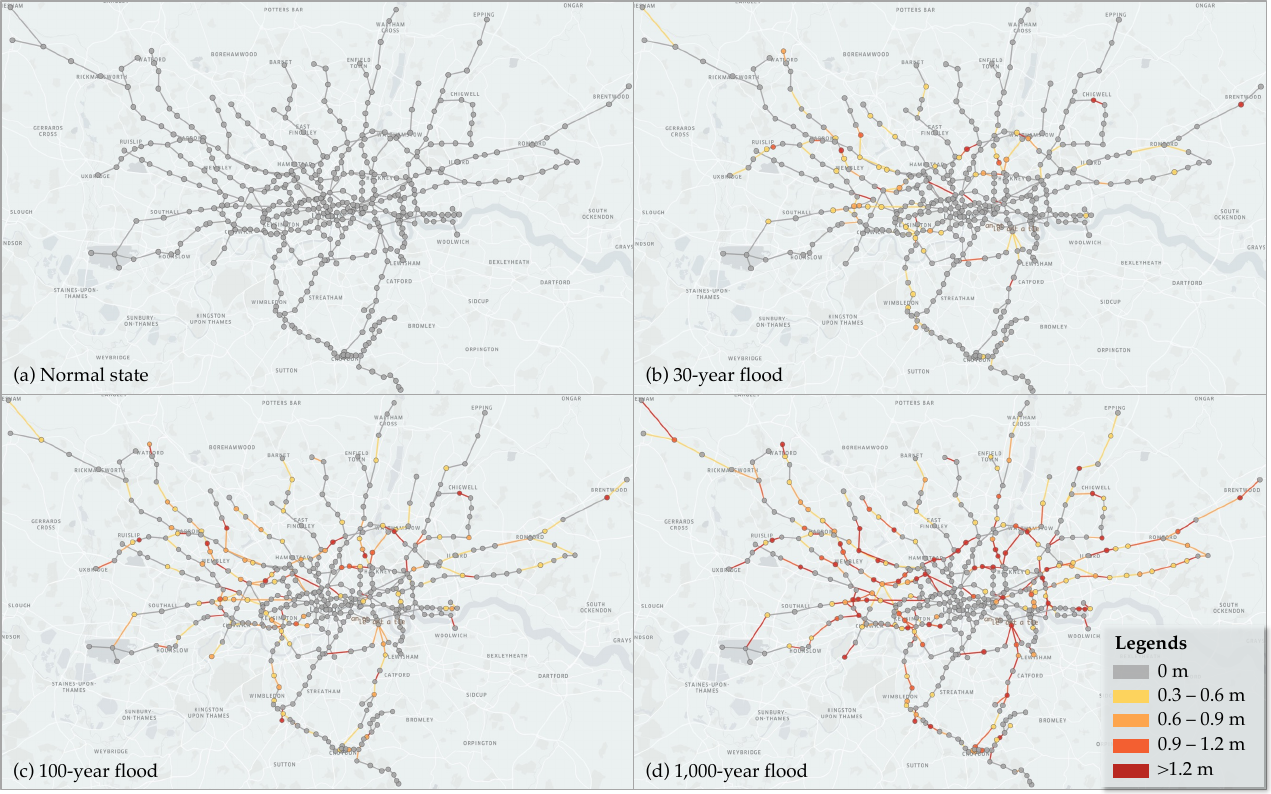}
    \caption {Flooded London URTS stations and track sections by inundation depth \citep{Bi2024Assessing}}
    \label{fig:B.1}
\end{figure}

\renewcommand{\thetable}{B.\arabic{table}} 
\setcounter{table}{0}

\begin{table}[ht!]
    \footnotesize
    \caption{Number of flooded London URTS elements (adapted from \cite{Bi2024Assessing})}
    \label{table:B.1}
    \renewcommand{\arraystretch}{1.2}
    \begin{tabular}{@{}p{0.18\linewidth}p{0.18\linewidth}p{0.18\linewidth}p{0.18\linewidth}p{0.18\linewidth}@{}}
    \hline
    {Element type} & {Flood depth} & \multicolumn{3}{c}{{Number of flooded elements in corresponding scenario}} \\
    \cmidrule(lr){3-5}
    & & {30-year flood} & {100-year flood} & {1,000-year flood} \\
    \hline
    Nodes & 0.3-0.6 m & 28 & 39 & 51 \\
          & 0.6-0.9 m & 9  & 27 & 27 \\
          & 0.9-1.2 m & 3  & 6  & 19 \\
          & $>$1.2 m  & 4  & 10 & 49 \\
          & Total    & 44 & 82 & 146 \\
    Edges & 0.3-0.6 m & 50 & 61 & 48 \\
          & 0.6-0.9 m & 15 & 43 & 41 \\
          & 0.9-1.2 m & 5  & 12 & 43 \\
          & $>$1.2 m  & 11 & 30 & 96 \\
          & Total    & 81 & 146 & 228 \\
    \hline
    \end{tabular}
\end{table}

%******************************** Appendix C ************************************

\newpage
\section{Disruption impacts on passenger travel (circuity factor = 1.2)}
\label{app:C}

\renewcommand{\thefigure}{C.\arabic{figure}}
\setcounter{figure}{0}

\begin{figure}[ht!] 
    \centering    
    \includegraphics[width=1.0\textwidth]{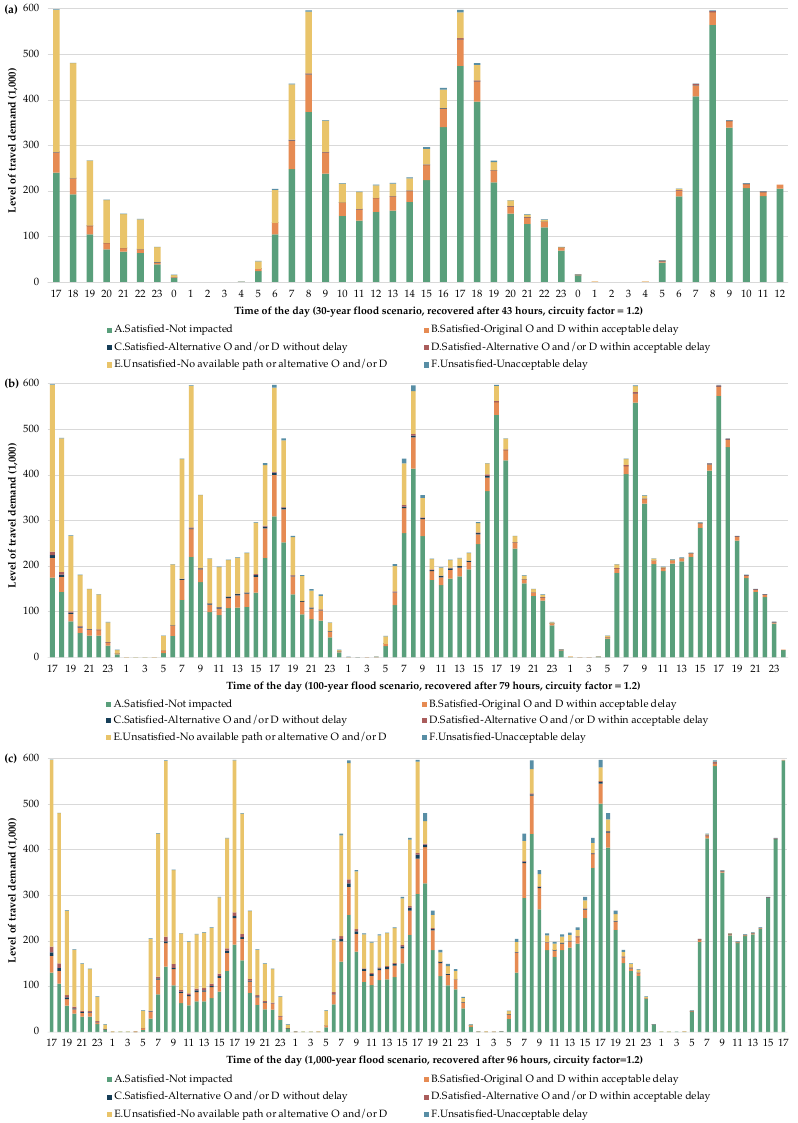}
    \caption {Hourly distribution of travel demand across impact categories (A)–(F) (circuity factor = 1.2)}
    \label{fig:C.1}
\end{figure}

%******************************** Appendix D ************************************

\newpage
\section{Hourly percentage of journeys classified into impact categories (A)–(F)}
\label{app:D}

\renewcommand{\thetable}{D.\arabic{table}} 
\setcounter{table}{0}

\begin{table}[ht!]
    \footnotesize
    \caption{Hourly percentage of journeys classified into impact categories under 30-year flood risk scenario}
    \label{table:D.1}
    \renewcommand{\arraystretch}{1.15}
    \begin{tabular}{c c c c c c c}
        \hline
        \multirow{2}{*}{Time} & \multicolumn{6}{c}{\begin{tabular}[c]{@{}c@{}}Percentage of journeys classified \\ into each impact category\end{tabular}} \\
        \cline{2-7}
         & A & B & C & D & E & F \\
        \hline
        Day 1, 17:00 & 40.3\% & 7.4\% & 0.1\% & 0.0\% & 52.1\% & 0.0\% \\
        18:00 & 40.3\% & 7.1\% & 0.1\% & 0.0\% & 52.5\% & 0.0\% \\
        19:00 & 39.4\% & 7.0\% & 0.1\% & 0.0\% & 53.4\% & 0.0\% \\
        20:00 & 40.0\% & 7.6\% & 0.1\% & 0.0\% & 52.2\% & 0.0\% \\
        21:00 & 44.8\% & 5.5\% & 0.1\% & 0.0\% & 49.6\% & 0.0\% \\
        22:00 & 47.1\% & 5.1\% & 0.1\% & 0.0\% & 47.6\% & 0.0\% \\
        23:00 & 51.0\% & 5.2\% & 0.1\% & 0.0\% & 43.7\% & 0.0\% \\
        0:00  & 60.7\% & 7.7\% & 0.1\% & 0.0\% & 30.2\% & 1.3\% \\
        1:00  & 53.6\% & 4.3\% & 0.0\% & 0.0\% & 42.1\% & 0.0\% \\
        2:00  & 88.0\% & 0.6\% & 0.0\% & 0.0\% & 11.4\% & 0.0\% \\
        3:00  & 88.9\% & 0.1\% & 0.0\% & 0.0\% & 11.0\% & 0.0\% \\
        4:00  & 34.2\% & 22.6\% & 0.0\% & 0.0\% & 43.2\% & 0.0\% \\
        Day 2, 5:00  & 52.1\% & 11.5\% & 0.0\% & 0.0\% & 35.2\% & 1.2\% \\
        6:00  & 51.1\% & 12.3\% & 0.0\% & 0.0\% & 35.6\% & 0.9\% \\
        7:00  & 56.9\% & 14.2\% & 0.1\% & 0.2\% & 28.2\% & 0.5\% \\
        8:00  & 62.8\% & 13.7\% & 0.1\% & 0.2\% & 22.9\% & 0.3\% \\
        9:00  & 67.0\% & 13.1\% & 0.1\% & 0.1\% & 19.3\% & 0.3\% \\
        10:00 & 67.4\% & 13.2\% & 0.1\% & 0.1\% & 18.9\% & 0.4\% \\
        11:00 & 68.1\% & 12.7\% & 0.1\% & 0.1\% & 18.6\% & 0.4\% \\
        12:00 & 72.0\% & 13.9\% & 0.1\% & 0.1\% & 13.0\% & 0.8\% \\
        13:00 & 72.0\% & 14.1\% & 0.1\% & 0.1\% & 12.9\% & 0.8\% \\
        14:00 & 76.9\% & 10.3\% & 0.1\% & 0.1\% & 11.8\% & 0.7\% \\
        15:00 & 75.7\% & 11.0\% & 0.1\% & 0.3\% & 12.0\% & 0.9\% \\
        16:00 & 79.8\% & 9.6\% & 0.0\% & 0.0\% & 9.9\% & 0.7\% \\
        17:00 & 79.5\% & 9.7\% & 0.0\% & 0.0\% & 10.1\% & 0.7\% \\
        18:00 & 82.5\% & 9.3\% & 0.0\% & 0.0\% & 7.6\% & 0.6\% \\
        19:00 & 82.3\% & 9.3\% & 0.0\% & 0.0\% & 7.7\% & 0.7\% \\
        20:00 & 83.4\% & 8.8\% & 0.0\% & 0.0\% & 7.3\% & 0.5\% \\
        21:00 & 85.9\% & 9.0\% & 0.0\% & 0.0\% & 4.7\% & 0.4\% \\
        22:00 & 87.7\% & 8.8\% & 0.0\% & 0.0\% & 3.1\% & 0.4\% \\
        23:00 & 89.0\% & 8.9\% & 0.0\% & 0.0\% & 2.1\% & 0.0\% \\
        0:00  & 93.4\% & 4.6\% & 0.0\% & 0.0\% & 2.1\% & 0.0\% \\
        1:00  & 81.0\% & 2.0\% & 0.0\% & 0.0\% & 17.0\% & 0.0\% \\
        2:00  & 93.0\% & 6.6\% & 0.0\% & 0.0\% & 0.4\% & 0.0\% \\
        3:00  & 91.3\% & 8.5\% & 0.0\% & 0.0\% & 0.1\% & 0.0\% \\
        4:00  & 91.6\% & 2.5\% & 0.0\% & 0.0\% & 5.8\% & 0.0\% \\
        Day 3, 5:00  & 91.8\% & 6.2\% & 0.0\% & 0.0\% & 2.0\% & 0.0\% \\
        6:00  & 92.1\% & 6.1\% & 0.0\% & 0.0\% & 1.9\% & 0.0\% \\
        7:00  & 93.5\% & 5.7\% & 0.0\% & 0.0\% & 0.9\% & 0.0\% \\
        8:00  & 94.7\% & 4.7\% & 0.0\% & 0.0\% & 0.6\% & 0.0\% \\
        9:00  & 95.5\% & 4.0\% & 0.0\% & 0.0\% & 0.5\% & 0.0\% \\
        10:00 & 95.3\% & 4.2\% & 0.0\% & 0.0\% & 0.5\% & 0.0\% \\
        11:00 & 95.3\% & 4.2\% & 0.0\% & 0.0\% & 0.5\% & 0.0\% \\
        12:00 & 96.0\% & 4.0\% & 0.0\% & 0.0\% & 0.0\% & 0.0\% \\
        \hline
    \end{tabular}%
\end{table}

\begin{table}[ht!]
    \footnotesize
    \caption{Hourly percentage of journeys classified into impact categories under 100-year flood risk scenario}
    \label{table:D.2}
    \renewcommand{\arraystretch}{1.15}
    \begin{minipage}[t]{0.48\textwidth}
    \centering
    \begin{tabular}{c c c c c c c c}
        \hline
        \multirow{2}{*}{Time} & \multicolumn{6}{c}{\begin{tabular}[c]{@{}c@{}}Percentage of journeys classified \\ into each impact category\end{tabular}} \\
        \cline{2-7}
         & A & B & C & D & E & F \\
        \hline
        Day 1, 17:00 & 29\% & 7\% & 1\% & 1\% & 61\% & 0\% \\
        18:00 & 30\% & 7\% & 1\% & 1\% & 61\% & 0\% \\
        19:00 & 30\% & 6\% & 1\% & 1\% & 62\% & 0\% \\
        20:00 & 30\% & 7\% & 1\% & 1\% & 62\% & 0\% \\
        21:00 & 32\% & 8\% & 1\% & 0\% & 58\% & 0\% \\
        22:00 & 35\% & 9\% & 0\% & 0\% & 56\% & 0\% \\
        23:00 & 34\% & 8\% & 0\% & 0\% & 57\% & 0\% \\
        0:00 & 39\% & 13\% & 1\% & 0\% & 47\% & 0\% \\
        1:00 & 26\% & 2\% & 0\% & 0\% & 72\% & 0\% \\
        2:00 & 86\% & 0\% & 0\% & 0\% & 13\% & 0\% \\
        3:00 & 88\% & 0\% & 0\% & 0\% & 11\% & 0\% \\
        4:00 & 8\% & 7\% & 0\% & 0\% & 85\% & 0\% \\
        Day 2, 5:00 & 19\% & 12\% & 0\% & 0\% & 68\% & 0\% \\
        6:00 & 23\% & 11\% & 0\% & 0\% & 66\% & 0\% \\
        7:00 & 29\% & 10\% & 0\% & 0\% & 60\% & 0\% \\
        8:00 & 37\% & 10\% & 0\% & 0\% & 52\% & 0\% \\
        9:00 & 46\% & 8\% & 1\% & 0\% & 45\% & 0\% \\
        10:00 & 46\% & 8\% & 1\% & 0\% & 45\% & 0\% \\
        11:00 & 46\% & 8\% & 1\% & 0\% & 45\% & 0\% \\
        12:00 & 50\% & 11\% & 1\% & 0\% & 37\% & 0\% \\
        13:00 & 50\% & 12\% & 1\% & 0\% & 36\% & 0\% \\
        14:00 & 48\% & 12\% & 1\% & 0\% & 37\% & 0\% \\
        15:00 & 48\% & 12\% & 1\% & 0\% & 38\% & 0\% \\
        16:00 & 51\% & 15\% & 1\% & 0\% & 31\% & 1\% \\
        17:00 & 52\% & 15\% & 1\% & 0\% & 31\% & 1\% \\
        18:00 & 52\% & 15\% & 1\% & 0\% & 31\% & 1\% \\
        19:00 & 52\% & 15\% & 1\% & 0\% & 31\% & 1\% \\
        20:00 & 52\% & 15\% & 1\% & 0\% & 31\% & 1\% \\
        21:00 & 56\% & 16\% & 1\% & 0\% & 25\% & 2\% \\
        22:00 & 58\% & 16\% & 0\% & 0\% & 23\% & 2\% \\
        23:00 & 57\% & 16\% & 1\% & 0\% & 24\% & 2\% \\
        0:00 & 69\% & 9\% & 1\% & 0\% & 20\% & 1\% \\
        1:00 & 57\% & 6\% & 0\% & 0\% & 36\% & 1\% \\
        2:00 & 89\% & 5\% & 0\% & 0\% & 4\% & 2\% \\
        3:00 & 89\% & 6\% & 0\% & 0\% & 2\% & 3\% \\
        4:00 & 31\% & 24\% & 0\% & 0\% & 44\% & 0\% \\
        Day 3, 5:00 & 52\% & 10\% & 0\% & 0\% & 35\% & 2\% \\
        6:00 & 56\% & 13\% & 0\% & 0\% & 28\% & 2\% \\
        7:00 & 62\% & 13\% & 1\% & 0\% & 22\% & 2\% \\
        8:00 & 69\% & 12\% & 0\% & 0\% & 17\% & 2\% \\
        \hline
    \end{tabular}
    \end{minipage}
    \hfill
    \begin{minipage}[t]{0.48\textwidth}
    \centering
    \begin{tabular}{c c c c c c c}
        \hline
        \multirow{2}{*}{Time} & \multicolumn{6}{c}{\begin{tabular}[c]{@{}c@{}}Percentage of journeys classified \\ into each impact category\end{tabular}} \\
        \cline{2-7}
         & A & B & C & D & E & F \\
        \hline
        (continued) 9:00 & 75\% & 11\% & 1\% & 0\% & 12\% & 1\% \\
        10:00 & 78\% & 9\% & 1\% & 0\% & 11\% & 0\% \\
        11:00 & 80\% & 9\% & 1\% & 0\% & 9\% & 0\% \\
        12:00 & 81\% & 9\% & 1\% & 0\% & 8\% & 0\% \\
        13:00 & 81\% & 9\% & 1\% & 0\% & 8\% & 0\% \\
        14:00 & 84\% & 7\% & 1\% & 0\% & 7\% & 0\% \\
        15:00 & 84\% & 7\% & 1\% & 0\% & 7\% & 0\% \\
        16:00 & 86\% & 7\% & 1\% & 0\% & 6\% & 0\% \\
        17:00 & 89\% & 5\% & 0\% & 0\% & 6\% & 0\% \\
        18:00 & 90\% & 5\% & 0\% & 0\% & 5\% & 0\% \\
        19:00 & 90\% & 5\% & 0\% & 0\% & 6\% & 0\% \\
        20:00 & 90\% & 4\% & 0\% & 0\% & 6\% & 0\% \\
        21:00 & 90\% & 5\% & 0\% & 0\% & 5\% & 0\% \\
        22:00 & 91\% & 5\% & 0\% & 0\% & 4\% & 0\% \\
        23:00 & 91\% & 5\% & 0\% & 0\% & 4\% & 0\% \\
        0:00 & 91\% & 4\% & 0\% & 0\% & 4\% & 0\% \\
        1:00 & 95\% & 1\% & 0\% & 0\% & 4\% & 0\% \\
        2:00 & 93\% & 7\% & 0\% & 0\% & 0\% & 0\% \\
        3:00 & 91\% & 9\% & 0\% & 0\% & 0\% & 0\% \\
        4:00 & 86\% & 6\% & 0\% & 0\% & 8\% & 0\% \\
        Day 4, 5:00 & 88\% & 4\% & 0\% & 0\% & 7\% & 0\% \\
        6:00 & 91\% & 4\% & 0\% & 0\% & 5\% & 0\% \\
        7:00 & 92\% & 4\% & 0\% & 0\% & 4\% & 0\% \\
        8:00 & 94\% & 3\% & 0\% & 0\% & 3\% & 0\% \\
        9:00 & 95\% & 3\% & 0\% & 0\% & 2\% & 0\% \\
        10:00 & 94\% & 3\% & 0\% & 0\% & 2\% & 0\% \\
        11:00 & 96\% & 3\% & 0\% & 0\% & 1\% & 0\% \\
        12:00 & 96\% & 3\% & 0\% & 0\% & 1\% & 0\% \\
        13:00 & 96\% & 3\% & 0\% & 0\% & 1\% & 0\% \\
        14:00 & 96\% & 3\% & 0\% & 0\% & 1\% & 0\% \\
        15:00 & 96\% & 3\% & 0\% & 0\% & 1\% & 0\% \\
        16:00 & 96\% & 3\% & 0\% & 0\% & 1\% & 0\% \\
        17:00 & 96\% & 3\% & 0\% & 0\% & 1\% & 0\% \\
        18:00 & 96\% & 3\% & 0\% & 0\% & 1\% & 0\% \\
        19:00 & 96\% & 3\% & 0\% & 0\% & 1\% & 0\% \\
        20:00 & 96\% & 3\% & 0\% & 0\% & 1\% & 0\% \\
        21:00 & 96\% & 3\% & 0\% & 0\% & 1\% & 0\% \\
        22:00 & 96\% & 4\% & 0\% & 0\% & 1\% & 0\% \\
        23:00 & 95\% & 4\% & 0\% & 0\% & 1\% & 0\% \\
        0:00 & 96\% & 4\% & 0\% & 0\% & 0\% & 0\% \\
        \hline
    \end{tabular}
\end{minipage}
\end{table}

\begin{table}[ht!]
    \footnotesize
    \caption{Hourly percentage of journeys classified into impact categories under 1,000-year flood risk scenario}
    \label{table:D.3}
    \renewcommand{\arraystretch}{1.15}
    \begin{minipage}[t]{0.48\textwidth}
    \centering
    \begin{tabular}{c c c c c c c c}
        \hline
        \multirow{2}{*}{Time} & \multicolumn{6}{c}{\begin{tabular}[c]{@{}c@{}}Percentage of journeys classified \\ into each impact category\end{tabular}} \\
        \cline{2-7}
         & A & B & C & D & E & F \\
        \hline
        Day 1, 17:00 & 22\% & 6\% & 1\% & 1\% & 70\% & 0\% \\
        18:00 & 22\% & 6\% & 1\% & 1\% & 70\% & 0\% \\
        19:00 & 22\% & 5\% & 1\% & 1\% & 71\% & 0\% \\
        20:00 & 22\% & 6\% & 1\% & 1\% & 70\% & 0\% \\
        21:00 & 22\% & 6\% & 1\% & 1\% & 70\% & 0\% \\
        22:00 & 24\% & 6\% & 1\% & 1\% & 68\% & 0\% \\
        23:00 & 22\% & 6\% & 1\% & 1\% & 71\% & 0\% \\
        0:00 & 27\% & 9\% & 1\% & 1\% & 63\% & 0\% \\
        1:00 & 14\% & 3\% & 0\% & 0\% & 84\% & 0\% \\
        2:00 & 86\% & 0\% & 0\% & 0\% & 13\% & 0\% \\
        3:00 & 88\% & 0\% & 0\% & 0\% & 11\% & 0\% \\
        4:00 & 2\% & 6\% & 0\% & 0\% & 92\% & 0\% \\
        Day 2, 5:00 & 10\% & 7\% & 0\% & 1\% & 82\% & 0\% \\
        6:00 & 14\% & 8\% & 0\% & 1\% & 77\% & 0\% \\
        7:00 & 19\% & 7\% & 1\% & 1\% & 72\% & 0\% \\
        8:00 & 24\% & 9\% & 1\% & 2\% & 65\% & 0\% \\
        9:00 & 29\% & 10\% & 1\% & 2\% & 58\% & 0\% \\
        10:00 & 29\% & 10\% & 1\% & 2\% & 57\% & 0\% \\
        11:00 & 30\% & 10\% & 1\% & 2\% & 57\% & 0\% \\
        12:00 & 31\% & 10\% & 1\% & 2\% & 56\% & 0\% \\
        13:00 & 31\% & 10\% & 1\% & 2\% & 56\% & 0\% \\
        14:00 & 32\% & 11\% & 1\% & 1\% & 54\% & 0\% \\
        15:00 & 30\% & 10\% & 1\% & 1\% & 57\% & 0\% \\
        16:00 & 31\% & 9\% & 1\% & 1\% & 58\% & 0\% \\
        17:00 & 32\% & 10\% & 1\% & 1\% & 57\% & 0\% \\
        18:00 & 33\% & 10\% & 1\% & 1\% & 56\% & 0\% \\
        19:00 & 32\% & 9\% & 1\% & 1\% & 57\% & 0\% \\
        20:00 & 33\% & 9\% & 1\% & 1\% & 56\% & 0\% \\
        21:00 & 34\% & 9\% & 1\% & 1\% & 55\% & 0\% \\
        22:00 & 36\% & 9\% & 1\% & 1\% & 54\% & 0\% \\
        23:00 & 34\% & 9\% & 1\% & 1\% & 56\% & 0\% \\
        0:00 & 42\% & 5\% & 1\% & 0\% & 52\% & 0\% \\
        1:00 & 16\% & 4\% & 0\% & 0\% & 79\% & 0\% \\
        2:00 & 87\% & 0\% & 0\% & 0\% & 13\% & 0\% \\
        3:00 & 89\% & 0\% & 0\% & 0\% & 11\% & 0\% \\
        4:00 & 3\% & 5\% & 0\% & 0\% & 92\% & 0\% \\
        Day 3, 5:00 & 21\% & 8\% & 0\% & 0\% & 71\% & 0\% \\
        6:00 & 29\% & 11\% & 1\% & 0\% & 58\% & 1\% \\
        7:00 & 35\% & 10\% & 1\% & 0\% & 52\% & 1\% \\
        8:00 & 43\% & 10\% & 1\% & 0\% & 44\% & 1\% \\
        9:00 & 50\% & 11\% & 1\% & 1\% & 37\% & 1\% \\
        10:00 & 51\% & 11\% & 2\% & 1\% & 35\% & 1\% \\
        11:00 & 52\% & 11\% & 2\% & 1\% & 34\% & 1\% \\
        12:00 & 53\% & 10\% & 2\% & 1\% & 33\% & 1\% \\
        13:00 & 53\% & 10\% & 2\% & 1\% & 34\% & 1\% \\
        14:00 & 52\% & 11\% & 2\% & 1\% & 34\% & 1\% \\
        15:00 & 51\% & 11\% & 2\% & 1\% & 35\% & 1\% \\
        16:00 & 50\% & 12\% & 1\% & 1\% & 35\% & 1\% \\
        17:00 & 51\% & 13\% & 1\% & 1\% & 34\% & 1\% \\
        \hline
    \end{tabular}
    \end{minipage}
    \hfill
    \begin{minipage}[t]{0.48\textwidth}
    \centering
    \begin{tabular}{c c c c c c c}
        \hline
        \multirow{2}{*}{Time} & \multicolumn{6}{c}{\begin{tabular}[c]{@{}c@{}}Percentage of journeys classified \\ into each impact category\end{tabular}} \\
        \cline{2-7}
         & A & B & C & D & E & F \\
        \hline
        (continued) 18:00 & 68\% & 17\% & 1\% & 0\% & 11\% & 3\% \\
        19:00 & 67\% & 16\% & 1\% & 0\% & 11\% & 3\% \\
        20:00 & 68\% & 16\% & 1\% & 0\% & 11\% & 3\% \\
        21:00 & 68\% & 15\% & 1\% & 1\% & 12\% & 2\% \\
        22:00 & 68\% & 15\% & 1\% & 0\% & 13\% & 2\% \\
        23:00 & 67\% & 16\% & 1\% & 0\% & 13\% & 2\% \\
        0:00 & 70\% & 15\% & 1\% & 0\% & 11\% & 2\% \\
        1:00 & 73\% & 9\% & 0\% & 0\% & 18\% & 1\% \\
        2:00 & 88\% & 8\% & 0\% & 0\% & 4\% & 0\% \\
        3:00 & 90\% & 8\% & 0\% & 0\% & 1\% & 0\% \\
        4:00 & 42\% & 15\% & 0\% & 0\% & 36\% & 7\% \\
        Day 4, 5:00 & 60\% & 22\% & 1\% & 0\% & 16\% & 1\% \\
        6:00 & 64\% & 21\% & 1\% & 0\% & 12\% & 2\% \\
        7:00 & 68\% & 17\% & 1\% & 0\% & 11\% & 3\% \\
        8:00 & 73\% & 14\% & 0\% & 0\% & 10\% & 3\% \\
        9:00 & 76\% & 13\% & 1\% & 0\% & 8\% & 2\% \\
        10:00 & 83\% & 7\% & 1\% & 0\% & 7\% & 2\% \\
        11:00 & 83\% & 7\% & 1\% & 0\% & 7\% & 2\% \\
        12:00 & 84\% & 7\% & 1\% & 0\% & 7\% & 2\% \\
        13:00 & 85\% & 6\% & 1\% & 0\% & 7\% & 2\% \\
        14:00 & 85\% & 6\% & 1\% & 0\% & 7\% & 2\% \\
        15:00 & 84\% & 6\% & 1\% & 0\% & 7\% & 2\% \\
        16:00 & 84\% & 7\% & 1\% & 0\% & 6\% & 2\% \\
        17:00 & 84\% & 7\% & 1\% & 0\% & 6\% & 2\% \\
        18:00 & 84\% & 7\% & 1\% & 0\% & 6\% & 2\% \\
        19:00 & 84\% & 7\% & 1\% & 0\% & 6\% & 2\% \\
        20:00 & 84\% & 7\% & 1\% & 0\% & 6\% & 2\% \\
        21:00 & 89\% & 4\% & 1\% & 0\% & 5\% & 0\% \\
        22:00 & 90\% & 4\% & 1\% & 0\% & 5\% & 0\% \\
        23:00 & 95\% & 1\% & 0\% & 0\% & 4\% & 0\% \\
        0:00 & 98\% & 0\% & 0\% & 0\% & 2\% & 0\% \\
        1:00 & 93\% & 0\% & 0\% & 0\% & 7\% & 0\% \\
        2:00 & 97\% & 0\% & 0\% & 0\% & 3\% & 0\% \\
        3:00 & 99\% & 0\% & 0\% & 0\% & 1\% & 0\% \\
        4:00 & 64\% & 1\% & 0\% & 0\% & 35\% & 0\% \\
        Day 5, 5:00 & 93\% & 1\% & 0\% & 0\% & 6\% & 0\% \\
        6:00 & 97\% & 2\% & 0\% & 0\% & 1\% & 0\% \\
        7:00 & 97\% & 1\% & 0\% & 0\% & 1\% & 0\% \\
        8:00 & 98\% & 1\% & 0\% & 0\% & 1\% & 0\% \\
        9:00 & 98\% & 1\% & 0\% & 0\% & 1\% & 0\% \\
        10:00 & 98\% & 1\% & 0\% & 0\% & 1\% & 0\% \\
        11:00 & 98\% & 1\% & 0\% & 0\% & 1\% & 0\% \\
        12:00 & 98\% & 1\% & 0\% & 0\% & 1\% & 0\% \\
        13:00 & 98\% & 1\% & 0\% & 0\% & 1\% & 0\% \\
        14:00 & 99\% & 1\% & 0\% & 0\% & 1\% & 0\% \\
        15:00 & 99\% & 0\% & 0\% & 0\% & 0\% & 0\% \\
        16:00 & 99\% & 0\% & 0\% & 0\% & 1\% & 0\% \\
        17:00 & 100\% & 0\% & 0\% & 0\% & 0\% & 0\% \\
         & & & & & & \\
        \hline
    \end{tabular}
\end{minipage}
\end{table}

%******************************** Bibliography ************************************
\clearpage

% Loading bibliography style file
\bibliographystyle{cas-model2-names}

% Loading bibliography database
\bibliography{cas-refs}

\end{document}